\documentclass[prd,twocolumn,floatfix,showpacs,preprintnumbers,nofootinbib]{revtex4}
\usepackage[utf8x]{inputenc}
\usepackage{mathrsfs}
\usepackage{amssymb}
\usepackage{amsmath}
\usepackage{graphicx}

\usepackage[caption = false]{subfig}
\usepackage[normalem]{ulem}
\usepackage{xcolor}
\definecolor{lcolor}{rgb}{0.5,0,0}
\definecolor{citcolor}{rgb}{0,0.3,0.0}

\usepackage[breaklinks,colorlinks,urlcolor=blue,citecolor=citcolor,linkcolor=lcolor]{hyperref}
\usepackage{mciteplus}

\newcommand{\rt}{{\mathbf{r}}}
\newcommand{\xt}{{\mathbf{x}}}
\newcommand{\bt}{{\mathbf{b}}}
\newcommand{\bti}{{\mathbf{b}_{i}}}
\newcommand{\yt}{{\mathbf{y}}}

\newcommand{\Deltat}{{\boldsymbol{\Delta}}}

\newcommand{\ud}{\, \mathrm{d}}

\newcommand{\tr}{\, \mathrm{Tr} \, }

\newcommand{\nc}{{N_\mathrm{c}}}

\newcommand{\gev}{\ \textrm{GeV}}
\newcommand{\fm}{\ \textrm{fm}}

\newcommand{\lqcd}{\Lambda_{\mathrm{QCD}}}
\newcommand{\as}{\alpha_{\mathrm{s}}}

\newcommand{\sigmap}{{ \sigma^\textrm{p}_\textrm{dip} }}

\newcommand{\dsigmap}{{\frac{\ud \sigma^\textrm{p}_\textrm{dip}}{\ud^2 \bt}}}

\newcommand{\xpom}{{x_\mathbb{P}}}

\newcommand{\der}{\mathrm{d}}

\newcommand{\A}{{\mathcal{A}}}

\begin{document}

\author{Heikki M\"antysaari}
\affiliation{
Physics Department, Brookhaven National Laboratory, Upton, NY 11973, USA
}

\author{Bj\"orn Schenke}
\affiliation{
Physics Department, Brookhaven National Laboratory, Upton, NY 11973, USA
}

\title{
Revealing proton shape fluctuations with incoherent diffraction at high energy
}

\pacs{24.85.+p, 13.60.-r}

\preprint{}

\begin{abstract}
The differential cross section of exclusive diffractive vector meson production in electron proton collisions carries important information on the geometric structure of the proton. More specifically, the coherent cross section as a function of the transferred transverse momentum is sensitive to the size of the proton, while the incoherent, or proton dissociative cross section is sensitive to fluctuations of the gluon distribution in coordinate space. We show that at high energies the experimentally measured coherent and incoherent cross sections for the production of $J/\Psi$ mesons are very well reproduced within the color glass condensate framework when strong geometric fluctuations of the gluon distribution in the proton are included. For $\rho$ meson production we also find reasonable agreement. We study in detail the dependence of our results on various model parameters, including the average proton shape, analyze the effect of saturation scale and color charge fluctuations and constrain the degree of geometric fluctuations. 
\end{abstract}

\maketitle

\section{Introduction}

Measuring the partonic structure of the proton has been one of the main motivations of deeply inelastic scattering (DIS) experiments. To date, the most precise data on the proton structure is provided by the H1 and ZEUS experiments at HERA~\cite{Aaron:2009aa,Abramowicz:2015mha}. Fundamentally, one is interested in the Wigner distributions of the proton's constituents \cite{Belitsky:2003nz}, which carry information on both the three dimensional momentum and spatial distributions. It is not known how to measure this distribution itself, such that 
typically some of the variables are integrated out. Most commonly both the spatial structure and transverse momentum are integrated out, yielding ordinary parton distribution functions (pdfs), only depending on the longitudinal momentum fraction $x$. 
If only either the transverse momentum or the spatial coordinates are integrated out, one obtains generalized parton distribution functions (GPDs) \cite{Mueller:1998fv,Ji:1996nm,Radyushkin:1997ki,Collins:1996fb} or transverse momentum dependent parton distribution functions (TMDs) \cite{Tangerman:1994eh,Mulders:2000sh,Kimber:2001sc,D'Alesio:2004up,Anselmino:2004ky,Anselmino:2005sh,Aybat:2011zv}, respectively. These pdfs also carry detailed information on the angular momentum carried by partons, including their spin and orbital motion. 

In this work we are interested in an additional piece of information, namely how much the spatial distribution of gluons within the proton fluctuates event-by-event. This information is experimentally accessible via exclusive incoherent diffractive vector meson production, namely scattering events that produce a single vector meson and, separated by a rapidity gap, remnants of the dissociated proton. Together with data on coherent diffractive vector meson production, in which the proton stays intact, the shape and fluctuations of the gluon distribution in the proton can be constrained \cite{Mantysaari:2016ykx}. 

We are particularly interested in large center of mass energies, where we are sensitive to the small $x$ part of the constituents in the proton. 
We will thus work in the framework of 
the color glass condensate (CGC) effective theory \cite{Iancu:2003xm,Gelis:2010nm} of quantum chromodynamics (QCD).
Within the CGC framework, experimental data on the proton structure function have been well reproduced \cite{Albacete:2010sy,Rezaeian:2012ji,Lappi:2013zma}. 
Furthermore, a large variety of observables in high energy collisions, including, for example, single~\cite{Tribedy:2010ab,Lappi:2013zma,Fujii:2013gxa,Ma:2014mri,Ducloue:2015gfa,Ducloue:2016pqr} and double inclusive~\cite{Albacete:2010pg,Stasto:2011ru,Lappi:2012nh} particle production in proton-proton and proton-nucleus collisions, can be described. The CGC framework has also been extensively applied to study diffractive DIS in current and future experiments, 
including ultra-peripheral heavy ion and proton-nucleus collisions~\cite{Kovchegov:1999ji,kuroda:2005by,Goncalves:2005yr,Kowalski:2008sa,Caldwell:2009ke,Lappi:2010dd,Toll:2012mb,Lappi:2013am,Lappi:2014foa}. Incoherent diffraction with proton targets, however, has only been studied in a few publications~\cite{Dominguez:2008aa,Marquet:2009vs}. The importance of significant geometric fluctuations in the description of the incoherent cross section measured at HERA was pointed out in a recent Letter~\cite{Mantysaari:2016ykx}, on which we expand in this work. In the near future, before the realization of an Electron Ion Collider~ \cite{Boer:2011fh,Accardi:2012qut}, new data in a wide range of photon-nucleon center-of-mass energies can be obtained from ultra-peripheral heavy ion~\cite{Abelev:2012ba,Abbas:2013oua,Khachatryan:2016qhq,ATLAS:2016vdy} and proton-nucleus collisions~\cite{TheALICE:2014dwa,CMS:2016nct}. 

Besides its fundamental interest, the fluctuating geometric shape of the proton is potentially an important ingredient for describing high multiplicity proton-proton and proton-nucleus collisions. Many collective phenomena have been observed in such collisions~\cite{Khachatryan:2010gv,Abelev:2012ola,Aad:2012gla,Adare:2013piz} (see also Ref.~\cite{Dusling:2015gta} for a recent review). One possible explanation for such collectivity are strong final state interactions, which are responsible for the generation of anisotropic flow in heavy ion collisions. They can be modeled by applying hydrodynamic simulations \cite{Gale:2013da}. A successful description of the experimental data in p+A collisions by hydrodynamic models with sophisticated initial states, requires knowledge of the proton's initial state geometry and its fluctuations \cite{Schenke:2014zha}. Various analyses of geometric and interaction strength (equivalent to proton size) fluctuations in p+p and p+A collisions have been performed in the literature \cite{Deng:2011at,Coleman-Smith:2013rla,Alvioli:2014eda,Aad:2015zza,Welsh:2016siu,Albacete:2016pmp}. To really test the physical picture of hydrodynamic behavior in small collision systems, it is necessary to constrain proton shape fluctuations from other data than that from p+A (or p+p) collisions themselves, which will be done in this work.

This paper is organized as follows. In Sec.\,\ref{sec:ddis} we present the calculation of diffractive vector meson production cross sections in the CGC framework. Various phenomenological corrections to the cross sections are analyzed in Sec.\,\ref{sec:corrections}. Geometric fluctuations are implemented in Sec.\,\ref{sec:fluctuations} and saturation scale fluctuations in Sec.\,\ref{sec:qsfluct}. The numerical results are shown in Sec.\,\ref{sec:results}. In the appendices we show the quantitative effect of the phenomenological corrections and study various parameter dependencies in more detail.

\section{Diffractive DIS in the dipole picture} \label{sec:ddis}
\begin{figure}[tb]
\centering
		\includegraphics[width=0.35\textwidth]{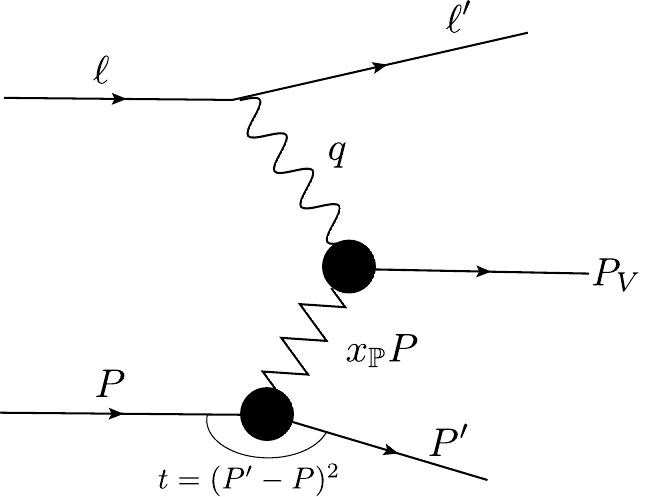} 
				\caption{Kinematics of diffractive deep inelastic scattering. The ``zigzag'' line represents pomeron exchange between the target and the photon. }
		\label{fig:ddis}
\end{figure}

We study the exclusive production of a vector meson $V$ with momentum $P_V$ in deeply inelastic scattering of leptons from protons:
\begin{equation}
	l(\ell) + p(P) \rightarrow l'(\ell') + p'(P') + V(P_V)\,.
\end{equation}
Here $\ell$ and $\ell'$ are the lepton momenta and $P$ and $P'$ are the proton momenta before and after the scattering, respectively.  The kinematics are illustrated in Fig.\,\ref{fig:ddis}. The Lorentz invariant quantities that characterize the scattering process are
\begin{align}
Q^2&\equiv-q^2=-(\ell-\ell')^2 \\
t &\equiv -(P'-P)^2 \\
\xpom &\equiv \frac{(P-P')\cdot q}{P \cdot q} = \frac{M^2+Q^2-t}{W^2+Q^2-m_N^2}.
\end{align}
Here $m_N$ is the proton mass, $M$ is the mass of the produced vector meson and $W^2=(P+q)^2$ is the total center-of-mass energy squared of the virtual photon-proton scattering. The fraction of the longitudinal momentum of the proton transferred to the vector meson is $\xpom$, where $\mathbb{P}$ stands for `pomeron'. The relation to the pomeron comes from the fact that in diffractive processes no color is exchanged between the proton and the produced system.  This means that there are no color strings between them, leading to a rapidity gap, i.e., a region in rapidity with no produced particles, which is used experimentally to identify diffractive events. The scattered proton $p'$ can either remain intact or break up, leading to coherent and incoherent diffractive events, respectively. 

In the Good-Walker picture~\cite{Good:1960ba}, diffraction is described in terms of states that diagonalize the scattering matrix. At high energy, these states are the ones where a virtual photon fluctuates into a quark-antiquark dipole with fixed transverse separation and impact parameter, and with a particular configuration of the target. The cross section is obtained by averaging over target configurations. 
Performing the average on the level of the scattering amplitude is equivalent to assuming that the target remains intact (\emph{coherent diffraction}), and the cross section is proportional to the average proton structure. On the other hand, averaging on the level of the cross section includes events in which the target breaks up, resulting in the total diffractive cross section. Subtracting the coherent contribution leaves us with only events where the target breaks up (\emph{incoherent diffraction}), which is proportional to the variance of the target profile, see e.g. Refs.\,\cite{Miettinen:1978jb,Frankfurt:1993qi,Frankfurt:2008vi,Caldwell:2009ke}.  For a pedagogical discussion of diffractive scattering and its description within perturbative QCD, we refer the reader to Ref.\,\cite{Barone:2002cv}.

Explicitly, in coherent diffraction the cross section can be written as~\cite{Miettinen:1978jb,Kowalski:2006hc}
\begin{equation}
\label{eq:coherent}
\frac{\der \sigma^{\gamma^* p \to V p}}{\der t} = \frac{1}{16\pi} \left| \langle \A^{\gamma^* p \to V p}(\xpom,Q^2,\boldsymbol{\Delta}) \rangle \right|^2,
\end{equation}
where $\A^{\gamma^* p \to V p}(\xpom,Q^2,\boldsymbol{\Delta})$ is the scattering amplitude.
The incoherent cross section can be written as the variance~\cite{Miettinen:1978jb} (see also e.g. Refs.~\cite{Frankfurt:1993qi,Frankfurt:2008vi,Caldwell:2009ke,Lappi:2010dd}):
\begin{align}\label{eq:incoherent}
\frac{\der \sigma^{\gamma^* p \to V p^*}}{\der t} = \frac{1}{16\pi} &\left( \left\langle \left| \A^{\gamma^* p \to V p}(\xpom,Q^2,\boldsymbol{\Delta})  \right|^2 \right\rangle \right. \notag\\ 
& ~~~ - \left. \left| \langle \A^{\gamma^* p \to V p}(\xpom,Q^2,\boldsymbol{\Delta}) \rangle \right|^2 \right)\,.
\end{align} 
We note that in \cite{Kovchegov:1999kx} and \cite{Kovner:2001vi} the different averaging procedures leading to above expressions are discussed in the context of a semi-classical description of small $x$ processes \cite{Buchmuller:1998jv}, such as the CGC picture employed in this work. In \cite{Kovner:2001vi} it is shown that the total diffractive cross section is obtained by averaging over the target fields on the level of the cross section, while in \cite{Kovchegov:1999kx} the coherent cross section is computed by averaging on the amplitude level as done in Eq.\,(\ref{eq:coherent}). 
\begin{figure}[tb]
\centering
		\includegraphics[width=0.35\textwidth]{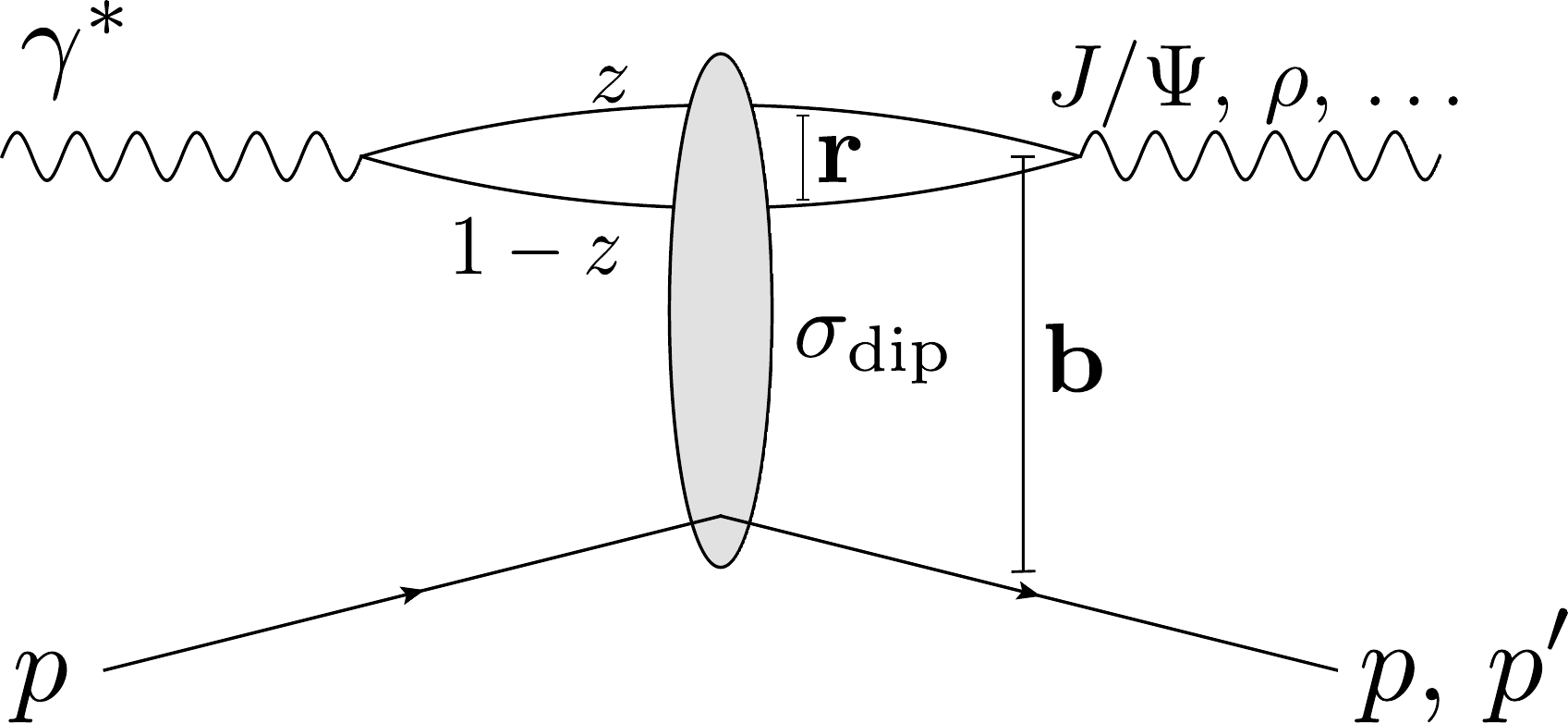} 
				\caption{Photon-proton scattering in the dipole picture. }
		\label{fig:photonproton}
\end{figure}
Following Ref.\,\cite{Kowalski:2006hc}, the scattering amplitude for diffractive vector meson production can be written as
\begin{multline}
\label{eq:diff_amp}
 \A^{\gamma^* p \to V p}_{T,L}(\xpom,Q^2, \boldsymbol{\Delta}) = i\int \der^2 \rt \int \der^2 \bt \int \frac{\der z}{4\pi}  \\ 
 \times (\Psi^*\Psi_V)_{T,L}(Q^2, \rt,z) \\
 \times e^{-i[\bt - (1-z)\rt]\cdot \boldsymbol{\Delta}}  \dsigmap(\bt,\rt,\xpom).
\end{multline}
Here the momentum transfer in the scattering process is $\Deltat=(P'-P)_\perp$, and throughout this work we assume $|\Deltat | \approx \sqrt{-t}$. The subscripts $T$ and $L$ refer to transverse and longitudinal polarization of the virtual photon.
Equation \eqref{eq:diff_amp} has a simple interpretation which is also illustrated in Fig.\,\ref{fig:photonproton}: First, an incoming virtual photon fluctuates into a quark-antiquark dipole with transverse size $\rt$ and $z$ being the longitudinal momentum fraction of the photon carried by the quark. This splitting is described by the virtual photon wave function $\Psi$, that can be calculated from perturbative QED (for a pedagogical discussion, see Ref.~\cite{Kovchegov:2012mbw}). The color dipole then scatters off the target proton with the dipole-proton cross section $\sigmap(\bt,\rt,\xpom)$, which we will discuss in detail below. This cross section is Fourier transformed into momentum space with the transverse momentum transfer $\Deltat$ being the Fourier conjugate to the center-of-mass of the dipole $\bt - (1-z)\rt$ (in the transverse plane and relative to the proton's center), where $\bt$ is the impact parameter~\cite{Kowalski:2006hc}. Finally, the scattered dipole forms the final state particle, in this case a vector meson with wave function $\Psi_V$. 

The vector meson wave function needs to be modeled. In this work we use the Boosted Gaussian wave function parametrization from Ref.~\cite{Kowalski:2006hc} as it  has been successfully used to describe HERA diffractive measurements.  There are also other wave functions available in the literature, but the different wave functions mainly affect the overall normalization of the results without significantly changing the $t$ dependence of the cross sections (see e.g.~\cite{Kowalski:2006hc}). Thus, our main results are not sensitive to the uncertainties related to the vector meson wave functions.

The dipole cross section $\sigmap$ is related to the forward elastic dipole-target scattering amplitude $N$ via the optical theorem as
\begin{equation}
\dsigmap(\bt,\rt,\xpom) = 2 N(\bt,\rt,\xpom).
\end{equation}
In the CGC framework the energy (or $\xpom$) evolution of the dipole amplitude is given by evolution equations that can be derived using perturbative techniques. Initial conditions for the small-$x$ evolution (dipole amplitude at initial Bjorken-$x$) can be determined by performing a fit to the HERA DIS data as in Refs.~\cite{Albacete:2010sy,Lappi:2013zma}. Then one can evolve the amplitude to smaller $x$ by solving the JIMWLK~\cite{JalilianMarian:1996xn,JalilianMarian:1997jx,JalilianMarian:1997gr,Iancu:2001md} or Balitsky-Kovchegov (BK) \cite{Balitsky:1995ub,Kovchegov:1999yj} evolution equation. 

Alternatively, the small-$x$ evolution can be modeled along with the impact parameter and $Q^2$ dependence of the dipole cross section, as done in the impact parameter dependent saturation (IPSat) model \cite{Kowalski:2003hm}. Because this approach has been very successful in describing a wide range of data from HERA and it avoids problems with the QCD evolution equations for finite size systems, such as the emergence of unphysical Coulomb tails \cite{GolecBiernat:2003ym,Schlichting:2014ipa}, we will use the IPSat model and the IP-Glasma model, \cite{Schenke:2012wb,Schenke:2013dpa}, where IPSat is coupled to classical Yang-Mills dynamics of the initial gluon fields.

In the IPsat model the dipole cross section is given by~\cite{Kowalski:2003hm}
\begin{equation}\label{eq:unfactbt}
\dsigmap(\bt,\rt,\xpom)
 = 2\,\left[ 1 - \exp\left(- {\mathbf r}^2  F(\xpom,\rt^2)\, T_p(\bt)\right) 
\right].
\end{equation} Here $T_p(\bt)$ is the proton (transverse) spatial profile function which is assumed to be Gaussian:
\begin{equation}\label{eq:Tp}
    T_p(\bt)=\frac{1}{2\pi B_p} e^{-\bt^2/(2B_p)}\,.
\end{equation}
The function $F$ is proportional to the 
DGLAP evolved gluon distribution~\cite{Bartels:2002cj},
\begin{equation}
F(\xpom, \rt^2) = 
\frac{ \pi^2 }{2 \nc} \as \left(\mu^2 \right) 
\xpom g\left(\xpom,\mu^2 \right),  
\label{eq:BEKW_F}
\end{equation}
with $\mu^2=\mu_0^2 + 4/\rt^2$. The proton width $B_p$, $\mu_0^2$ and the initial condition for the DGLAP evolution of the gluon distribution $\xpom g$ are parameters of the model. They are obtained in Ref.~\cite{Rezaeian:2012ji} by performing fits to HERA DIS data. For consistency with these fits we shall use the same scale $\mu^2$ also in the calculation of the diffractive cross section. See however Ref.~\cite{Gotsman:2001ne} for a discussion of a possible $|t|$ dependence of the scale choice in diffractive scattering. We use a charm mass of $m_c=1.4 \gev$.

In the IP-Glasma model \cite{Schenke:2012wb} the dipole amplitude $N$ can be calculated from the Wilson lines $V(\xt)$ as
\begin{equation}\label{eq:dipoleipglasma}
	N\left( \bt = \frac{\xt + \yt}{2}, \rt = \xt - \yt , \xpom  \right) = 1 - \frac{1}{\nc} \tr \left( V(\xt) V^\dagger(\yt) \right).
\end{equation}
Here the $\xpom$ dependence of the Wilson lines is left implicit. 
To get the Wilson lines, we first sample the color charges $\rho^a(\xt)$ from a Gaussian distribution
\begin{equation}
\langle \rho^a(x^-, \xt) \rho^b(y^-,\yt) \rangle = g^2 \delta^{ab} \delta^{(2)}(\xt-\yt) \delta(x^- - y^-) \mu^2.
\end{equation}
The color charge density  $g\mu$ is set to be proportional to the saturation scale $Q_s(\xpom, \xt)$ determined from the IPsat model. 
We treat the proportionality constant as a free parameter that mainly affects the overall normalization of our results. We will use $Q_s = 0.7 g^2 \mu$ when we include geometric fluctuations of the proton and $Q_s=0.65 g^2 \mu$ without. For a more detailed discussion on the relation between the saturation scale and color charge density, we refer the reader to Ref.~\cite{Lappi:2007ku}.

Solving the Yang-Mills equations for the gluon fields, one obtains
 \begin{equation}\label{eq:wilson}
  V (\xt) = P \exp\left({-ig\int dx^{-} \frac{\rho(x^-,\xt)}{\boldsymbol{\nabla}^2+m^2} }\right)\,.
\end{equation}
Here $P$ indicates path ordering and $m$ is an infrared cutoff. Its role is to suppress infrared long-distance Coulomb tails, and consequently it affects the proton size. Generally one expects $m \sim \lqcd$, and unless otherwise noted we will use $m=0.4\gev$. 
Sensitivity on the infrared cutoff $m$ is discussed in Appendix~\ref{app:m}.

The path ordering is calculated by discretizing the expression in (\ref{eq:wilson}) as
\begin{equation}
V(\xt) = \prod_{k=1}^{N_y} \exp \left({-ig \frac{\rho_k(\xt)}{\boldsymbol{\nabla}^2+m^2} }\right)\,.
\end{equation}
This corresponds to dividing the longitudinal direction into $N_y$ slices. The continuum limit is obtained by taking $N_y \to \infty$. In our calculations we use $N_y=100$. We have checked that for $N_y>100$ our results remain unchanged.

Calculations are performed on a 2-dimensional lattice with transverse spacing $a=0.02\,{\rm fm}$. We have checked that smaller lattice spacings do not alter the results. For more details on the IP-Glasma framework, the reader is referred to Ref.~\cite{Schenke:2012fw}.

As already discussed above, the coherent diffractive cross section is related to the Fourier transform of the dipole cross section $\sigmap$ from coordinate space to momentum space (see Eqs.\,\eqref{eq:coherent} and \eqref{eq:diff_amp}). Thus, the coherent cross section as a function of $|t|$ is directly related to the Fourier transform of the impact parameter profile of the proton. In the IPsat model the density profile is Gaussian, resulting in an approximately Gaussian spectrum in momentum space.
The proton size can then be characterized by the diffractive slope $B_D$ defined by fitting the coherent cross section by a function $\sim e^{-B_D |t|}$ in the small $|t|$ region. Notice that $B_D$ is not exactly the $B_p$ parameter in the IPsat model. The growth of the proton size (parameter $B_D$) as a function of energy has been observed at HERA~\cite{Aktas:2005xu} and in ultra-peripheral collisions by the ALICE collaboration~\cite{TheALICE:2014dwa}. Because in the IPsat model the density profile is assumed to factorize from the gluon distribution function $xg$, it is not possible to explain this measured proton growth within this framework as the width of the Gaussian does not change when the overall normalization (gluon density) increases~\cite{Kowalski:2006hc}. When performing explicit small $x$ QCD evolution as done in \cite{Schlichting:2014ipa} the growth of the proton with energy naturally emerges. 

The incoherent cross section, on the other hand, is given by the variance of the scattering amplitude (see Eq.~\eqref{eq:incoherent}). Thus, it is proportional to the event-by-event fluctuations of the proton density profile in coordinate space. As discussed in Ref.~\cite{Miettinen:1978jb}, at small $|t|$ it is dominated by fluctuations of the overall proton density (in our case driven by the value of $Q_s$ and possible color charge fluctuations). As we will demonstrate, at larger $|t|$, the effect of these fluctuations is negligible compared to the contribution originating from the geometric fluctuations.

\section{Phenomenological corrections}
\label{sec:corrections}
\subsection{Real part of the diffractive amplitude}
Derivation of the diffractive scattering amplitude \eqref{eq:diff_amp} relies on an assumption that the dipole scattering amplitude is purely real and the diffractive amplitude  imaginary. The real part of the amplitude can be taken into account by multiplying the calculated cross section by a factor 
 $(1+\beta^2)$, where the ratio of real to imaginary parts of the scattering amplitude is~\cite{Kowalski:2006hc}
\begin{equation}
\label{eq:realpart_beta}
\beta = \tan \frac{\pi \lambda}{2},
\end{equation}
where 
 \begin{equation}
 \lambda = \frac{\der \ln \A_{T,L}^{\gamma^* p \to V p} }{\der \ln 1/\xpom}.
 \end{equation}
In our calculation this correction is calculated without any event-by-event fluctuations.

 Because in the IP-Glasma framework the dipole amplitude has both real and imaginary parts, we do not include the real part correction when an IP-Glasma proton is used. However, we note that the contribution from the imaginary part of the dipole amplitude to the cross section is around 1\%, significantly less than the  correction $\sim 10\%$ calculated from Eq.~\eqref{eq:realpart_beta} in the kinematics relevant to this work (see Appendix\,\ref{appendix:corrections}).
 
\subsection{Skewedness correction}
 At lowest order the dipole-target scattering involves an exchange of two gluons, because there cannot be an exchange of color charge.
 The two gluons in the target are probed at different values of Bjorken $x$ ($x_1$ and $x_2$ satisfying $x_1 - x_2 = \xpom$). Because we calculate the imaginary part of the scattering amplitude, the dominant contribution is obtained when the intermediate propagators are close to the mass shell. Thus, the first gluon exchange has to bring the $q \bar q$ dipole mass close to the mass of the produced vector meson. Then, there is only a significantly smaller longitudinal momentum fraction $x_2$ left for the second gluon. The dominant kinematical regime is then $x_2 \ll x_1 \approx \xpom$~\cite{Martin:1997wy,Shuvaev:1999ce, Martin:1999wb}.

In the IPsat model the collinear factorization gluon distribution $\xpom g(\xpom, \mu^2)$ is corrected to correspond to the off-diagonal (or skewed) distribution, which depends on both $x_1$ and $x_2$, by multiplying it by a skewedness factor $R_g$ following the prescription of Ref.~\cite{Kowalski:2006hc}:
 \begin{equation}
 \label{eq:skew}
 R_g = 2^{2\lambda_g+3} \frac{\Gamma(\lambda_g+5/2)}{\sqrt{\pi}\,\Gamma(\lambda_g+4)}
 \end{equation}
 with 
 \begin{equation}
 \lambda_g = \frac{\der \ln \xpom g(\xpom,\mu^2)}{\der \ln 1/\xpom}.
 \end{equation}
 In the IP-Glasma model the gluon distribution function does not enter explicitly in the calculation of the diffractive scattering amplitude. In that case the skewedness correction is approximated by calculating its effect to the diffractive cross section within the IPsat model without geometrical fluctuations, and using the obtained correction factors to scale the calculated diffractive cross section. 

Especially the skewedness correction is numerically important and needed to describe the HERA diffractive measurements. We will study the relative importance of these corrections in Appendix\,\ref{appendix:corrections}.

\section{Fluctuating proton shape}
\label{sec:fluctuations}
While the average (or root-mean-square) proton radius\footnote{One can define e.g. the magnetic, charge \cite{RevModPhys.28.214,Bernauer:2010wm,Zhan:2011ji,Antognini417}, Zemach \cite{Friar:2003zg,Distler:2010zq}, axial \cite{Meissner:1987ge} and gluonic \cite{Kawasaki:1998gn,Caldwell:2010zza} radius of the proton. In this work we deal with the gluonic content of the proton.} is constrained relatively well, little is known about fluctuations in the proton's geometry. Here we explore several models for the fluctuating shape of the proton's gluon distribution and use experimental data on incoherent diffractive vector meson production to constrain the degree of fluctuations.

\subsection{Constituent quark proton}

\label{sec:constituentquark}

\begin{figure}
	\subfloat[$B_{qc}=3.3\gev^{-2}, B_q=0.7\gev^{-2}$]{\includegraphics[width=0.4\textwidth]{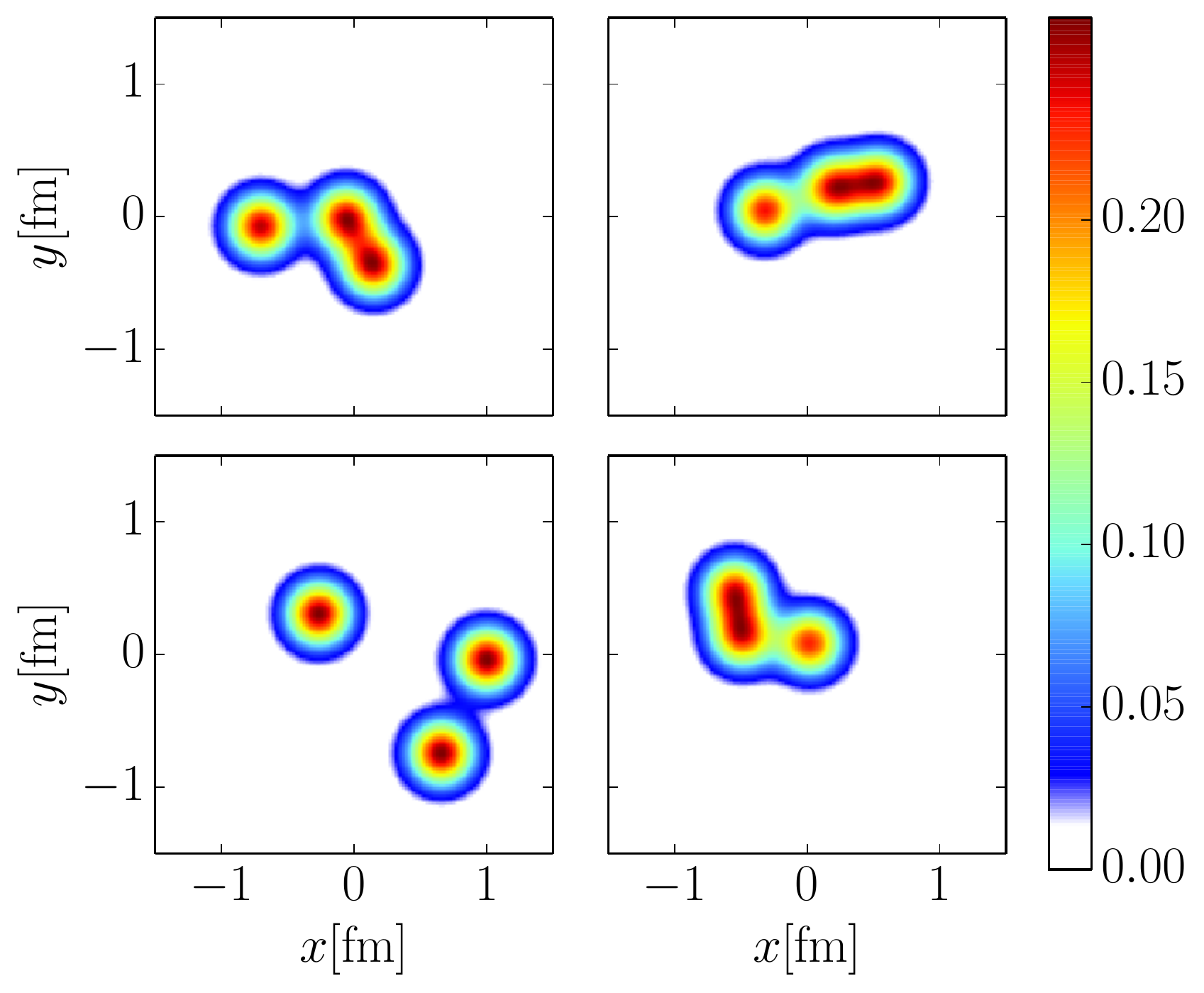}}
	\hfill
	\subfloat[$B_{qc}=1.0\gev^{-2}, B_q=3.0\gev^{-2}$]{\includegraphics[width=0.4\textwidth]{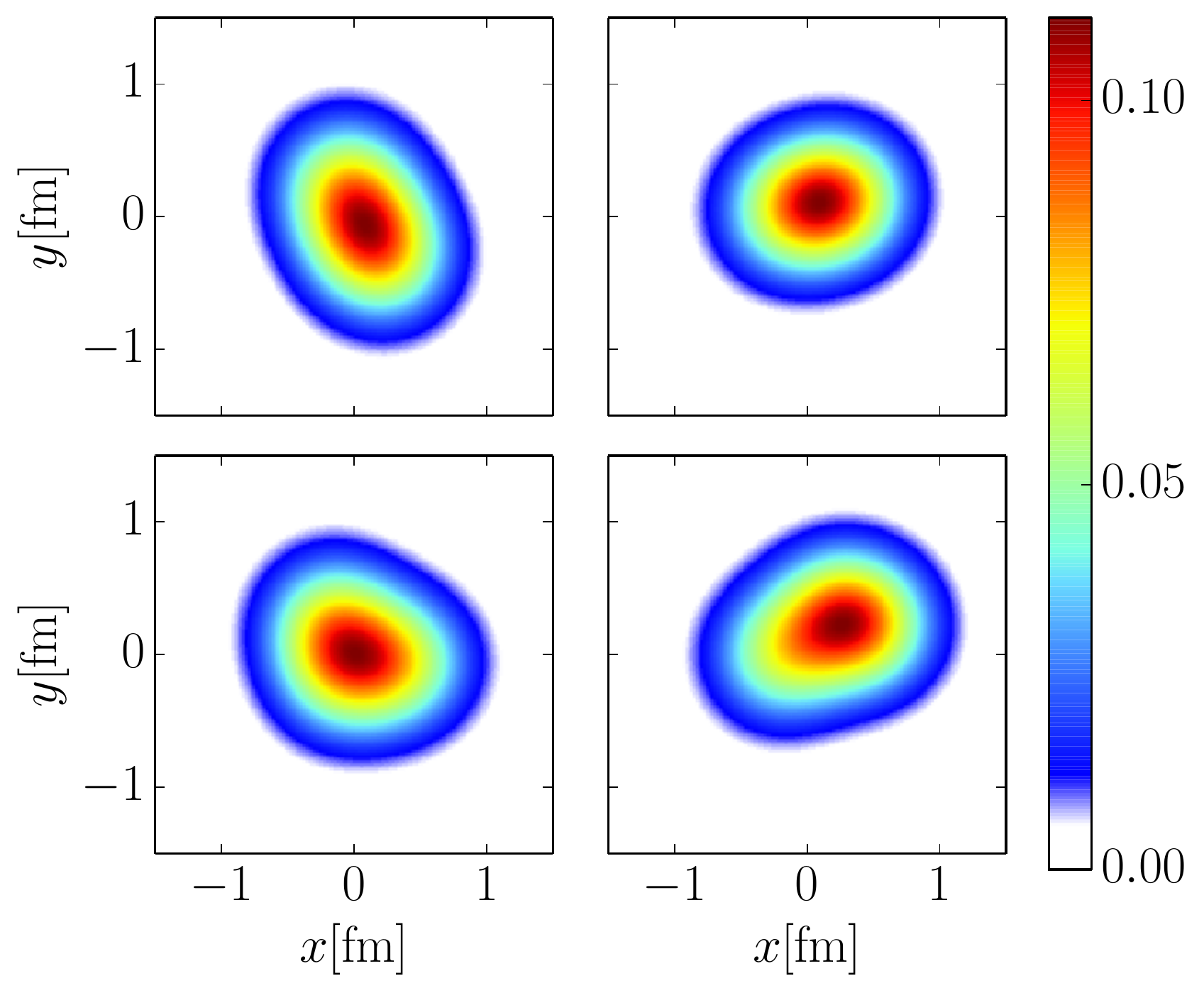}}
	\caption{Examples of proton density profiles at $x\approx 10^{-3}$ with two parametrizations used in this work.}
	\label{fig:ipsat_constituentquark_protons}
\end{figure}

The simplest profile we use to model proton event-by-event fluctuations is inspired by the constituent quark picture. Here, the large-$x$ valence quarks can be thought of as sources of small-$x$ gluons, emitted around the constituent quarks \cite{Schlichting:2014ipa}. 

We implement this picture by sampling the constituent quarks' positions in the transverse plane relative to the origin, $\bti$, from a  Gaussian distribution with width $B_{qc}$. The angular distribution of quarks is assumed to be uniform and we neglect any possible correlations between the quark positions. 
The density profile of each constituent quark in the transverse plane is also assumed to be Gaussian
\begin{equation}
T_q(\bt) = \frac{1}{2\pi B_q} e^{-\bt^2/(2B_q)}\,,
\end{equation}
with width parameter $B_q$.
This corresponds to the replacement
\begin{equation}\label{eq:TqReplace}
T_p(\bt) \rightarrow \frac{1}{N_q} \sum_{i=1}^{N_q} T_q(\bt-\bti)
\end{equation}
in Eq.~\eqref{eq:unfactbt}. $N_q$ can be interpreted as the number of large $x$ partons, typically chosen to be 3, for the three constituent quarks. We will also study larger values of $N_q$ in Appendix\,\ref{appendix:nq}, representing the situation of additional large $x$ gluons or sea-quarks.

For fixed $N_q$, the degree of fluctuations is controlled by the parameters $B_{qc}$ and $B_{q}$.  Examples of the sampled proton density profiles for $N_q=3$ are given in Fig.\,\ref{fig:ipsat_constituentquark_protons}. We show a ``lumpy'' proton configuration in panel a) and a ``smooth'' proton that has little fluctuations in panel b). In case of no geometric fluctuations, when the proton density profile is Gaussian with width $B_p$ (see Eq.~\eqref{eq:Tp}), the two-dimensional  gluonic root mean square radius of the proton is $r_p=\sqrt{2B_p}$. When coherent HERA data is fitted, one obtains $r_p=0.55 \fm$. Similarly, we can define the average radius of our fluctuating proton to be $\sqrt{2(B_q + B_{qc})}$, which in case of the parameter sets used in Fig.\,\ref{fig:ipsat_constituentquark_protons} has the same value.

\begin{figure}[htb]
  \centering
  \includegraphics[width=0.5\textwidth]{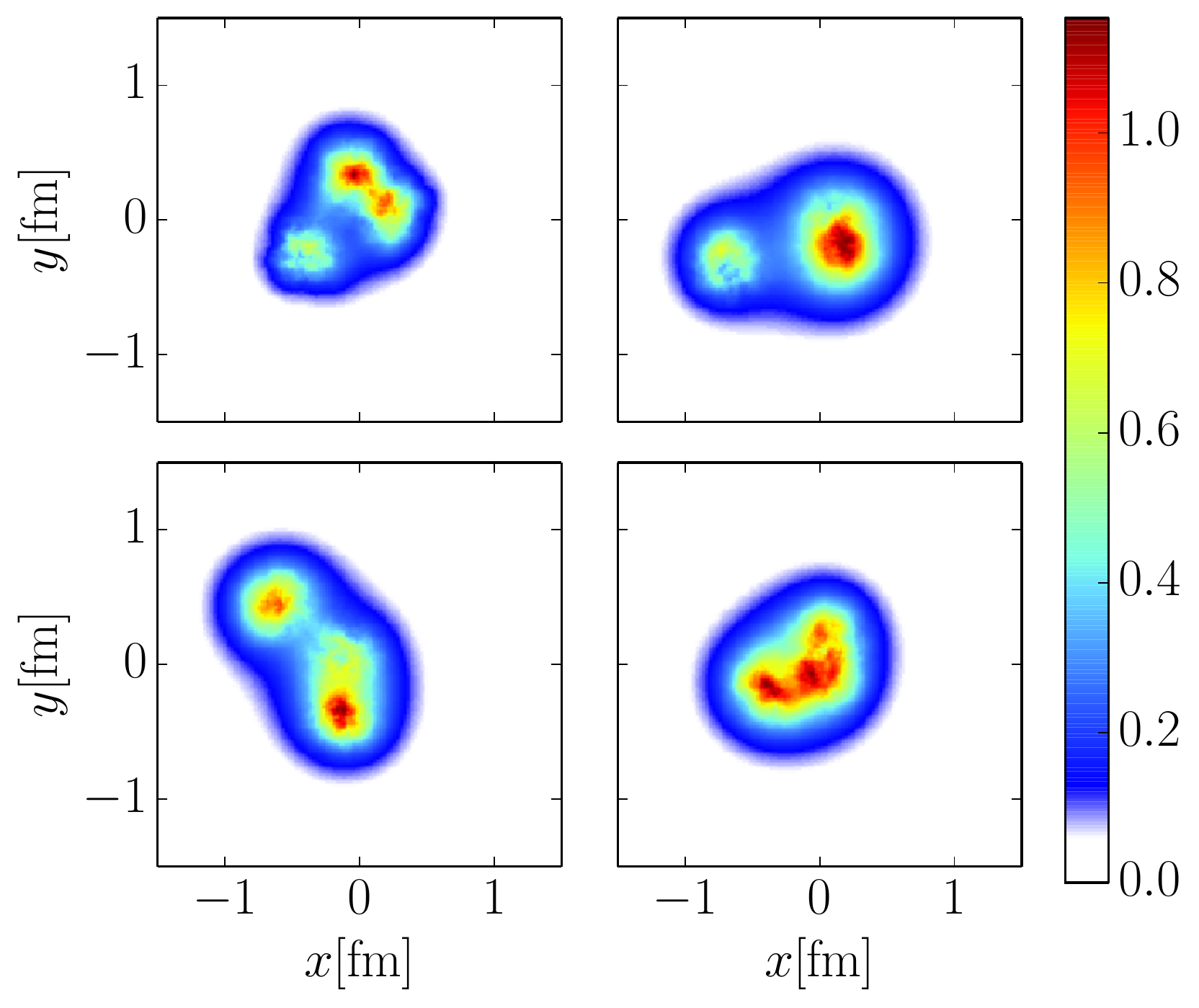} 
  \caption{Illustration of the proton density profile ($1.0 - \mathrm{Re} \tr V(x,y)/\nc$) obtained from the IP-Glasma framework at $x\approx 10^{-3}$ with parameters $B_{qc}=3.0\gev^{-2}, B_q=0.3\gev^{-2}$ and $m=0.4 \gev$.}
  \label{fig:ipglasma}
\end{figure}

In the IP-Glasma model geometric fluctuations are implemented by first performing the replacement \eqref{eq:TqReplace} in the IPsat model, which then provides the saturation scale values according to the modified thickness functions. In the IP-Glasma framework the additional parameter $m$ controls the infrared physics and thus affects the spatial size of the gluon distribution. Because of this the values for the parameters $B_{qc}$ and $B_{q}$ in both models cannot be directly compared. Examples of the proton density profiles obtained from the IP-Glasma model with the parametrization used in this work are illustrated in Fig.\,\ref{fig:ipglasma} by showing $1-\mathrm{Re}  \tr V(\xt)/\nc$.

The total photon-proton cross section, and the proton structure functions, are proportional to the integral of the dipole amplitude over impact parameter. As the modification \eqref{eq:TqReplace} is done in the exponent and the impact parameter dependence factorizes only in the dilute region, the replacement \eqref{eq:TqReplace} affects the overall normalization of, for example, $F_2$.
 In practice, including geometric fluctuations  ($B_{qc}=3.3\gev^{-2}, B_q=0.7\gev^{-2}$) decreases $F_2$ at $x\sim 10^{-3}, Q^2\sim 10\gev^2$ by approximately $8\%$. The diffractive cross section changes more, as it is proportional to the squared amplitude. Ideally one should perform a new fit to HERA DIS data with geometric fluctuations included, but this is beyond the scope of this work.
However, this normalization uncertainty is similar for both coherent and incoherent cross sections and will not affect our conclusions about the required amount of geometric fluctuations in the proton wave function.

To determine the sensitivity on the details of the assumed proton shape we will also calculate the diffractive cross sections using a three-dimensional exponential  density profile for the constituent quark
\begin{equation}
\label{eq:exp}
 T_q(b) = \frac{1}{8\pi  \tilde B_q^3 } e^{-b/  \tilde B_q }\,,
\end{equation}
and sample the constituent quark locations from a three-dimensional exponential distribution $\sim e^{-b/\tilde B_{qc}}$. The sampled quarks are then projected on the transverse plane. We note that the resulting transverse density profile is not exactly exponential.

\subsection{Stringy proton }
\label{sec:fluxtube}
In order to explore the dependence on the model details we also implement the geometric fluctuations using a color string inspired picture. Here, the idea is that based on quenched lattice QCD calculations, the constituent quarks are connected via gluon fields that merge at the Fermat point\footnote{The Fermat point of a triangle is defined such that the total distance from that point to the vertices of the triangle is the smallest possible.} of the quark triangle~\cite{Bissey:2006bz} (see also Ref.~\cite{Coleman-Smith:2013rla}). We are not aware of calculations beyond the quenched approximation, which would be a more appropriate input to our model.

We implement this picture by sampling the constituent quark positions from a three dimensional Gaussian distribution with width $B_{t}$. Then, the density profile is obtained by connecting the constituent quarks to the Fermat point of the triangle by tubes whose transverse shape is Gaussian with width $B_r$. The 2-dimensional density profile of the proton $T_p(\bt)$ is then obtained by integrating over the longitudinal direction. 

\begin{figure}[htb]
  \centering
  \includegraphics[width=0.5\textwidth]{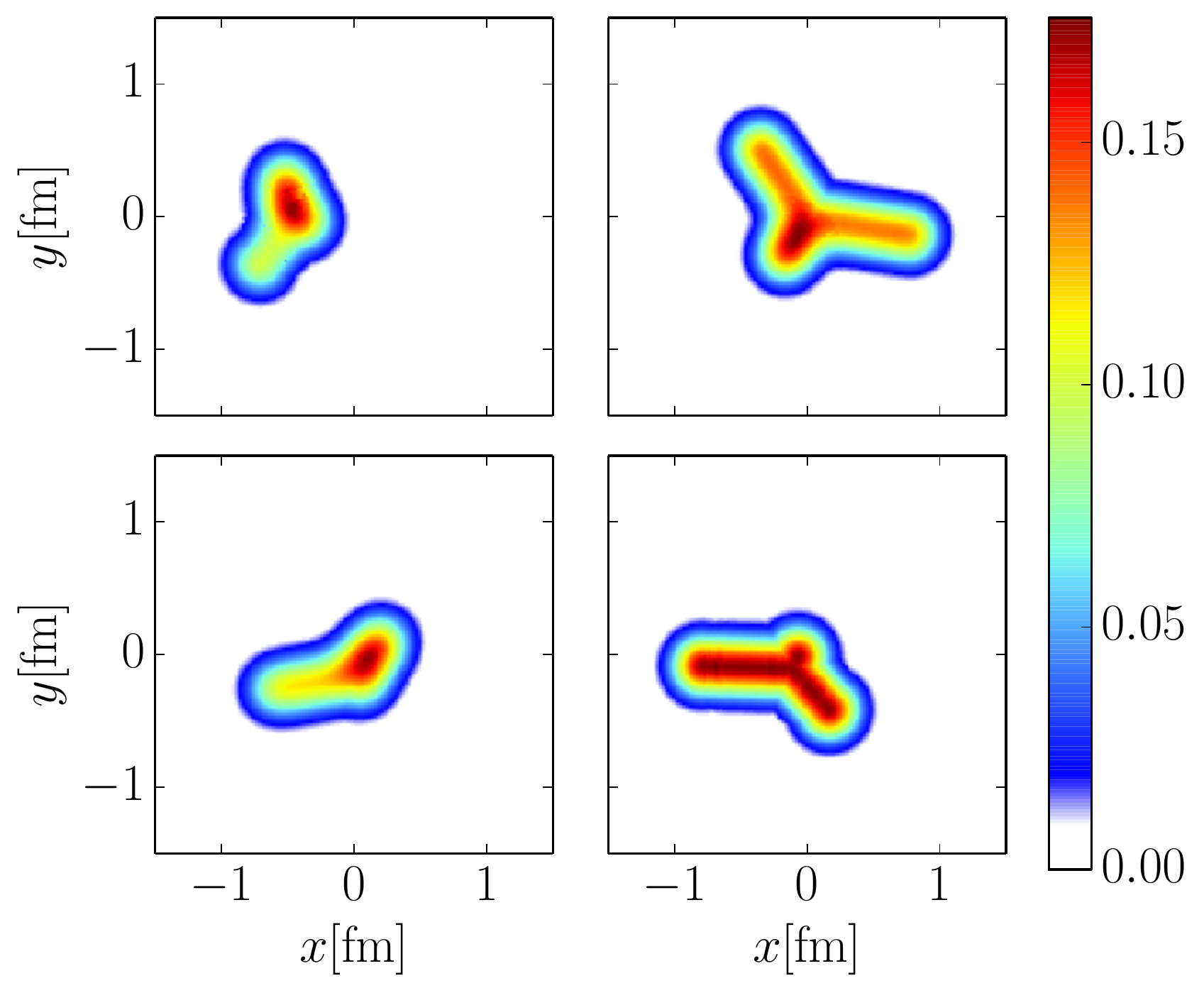} 
  \caption{Example density profiles of the ``stringy proton'' in the transverse plane at $x\approx 10^{-3}$ with parameters $B_{t}=4.2\gev^{-2}, B_r=0.6\gev^{-2}$}
  \label{fig:fluxtube}
\end{figure}

In this picture the total gluonic content of the proton also fluctuates event-by-event, as when the quarks are sampled to be further away from each other, the flux tubes are longer at a constant density, leading to more gluons in the proton. This adds normalization fluctuations to the picture, which are similar to those introduced by saturation scale fluctuations (see the following section). The overall normalization factor, which controls the energy density of the tube, is fixed by requiring that the proton structure function $F_2$ calculated from the stringy proton at $Q^2=10\gev^2$, $x=10^{-3}$ is the same as 
that from the original IPsat parametrization without fluctuations. Example density profiles (integrated over the longitudinal direction) are shown in Fig.\,\ref{fig:fluxtube}. The parameters $B_{t}$ and $B_r$ are fixed by requiring a good description of HERA coherent and incoherent diffractive $J/\Psi$ production measurements~\cite{Alexa:2013xxa}.

\section{Saturation scale fluctuations}
\label{sec:qsfluct}

Experimentally observed multiplicity distributions and rapidity correlations in p+p collisions can be explained in the IP-Glasma framework when
the saturation scale fluctuates according to \cite{McLerran:2015qxa,Bzdak:2015eii}
\begin{equation}
\label{eq:qsfluct}
P( \ln Q_s^2 / \langle Q_s^2 \rangle ) = \frac{1}{\sqrt{2\pi}\sigma} \exp \left[- \frac{\ln^2 Q_s^2/\langle Q_s^2\rangle}{2\sigma^2}\right],
\end{equation}
with the amount of fluctuations controlled by $\sigma \sim 0.5$. 

Because the log-normal distribution \eqref{eq:qsfluct} leads to the expectation value $E[Q_s^2 / \langle Q_s^2 \rangle]=e^{\sigma^2/2}$, sampling the $Q_s$ fluctuations directly from that distribution would make the average $Q_s^2$ to be $\approx 13\%$ larger (for $\sigma=0.5$) than in case of no saturation scale fluctuations. This would not be consistent with the IPsat model fit to the HERA data. Thus, when the saturation scale is sampled from the distribution \eqref{eq:qsfluct}, we normalize it by the mean of the distribution in order to get a fluctuating $Q_s$ distribution that always results in positive saturation scales and does not change the desired mean value.

In our constituent quark picture a natural way to include $Q_s$ fluctuations is to let the saturation scale of each constituent quark fluctuate independently. In case of no geometric fluctuations, we implement the $Q_s$ fluctuations 
by dividing the transverse space into a grid, where the cell size is set by the typical $1/Q_s^2$ (cf. \cite{Dumitru:2012yr}), which for the EIC and HERA kinematics we consider corresponds to $a\times a$ cells with $a\sim 0.4 \fm$.

\section{Results}
\label{sec:results}
We present results on coherent and incoherent diffractive vector meson production from the IPsat model with and without geometric fluctuations in 
Section \ref{sec:ipsat}. We show the effect of saturation scale fluctuations in Section \ref{sec:Qsfluc} and present results for the same observables in the IP-Glasma model in Section \ref{sec:ipglasma}.

\subsection{IPsat}\label{sec:ipsat}

We start by calculating the diffractive $J/\Psi$ photoproduction ($Q^2=0$) cross section that has been measured at HERA~\cite{Aktas:2005xu,Chekanov:2002xi,Chekanov:2002rm,Aktas:2003zi,Alexa:2013xxa} in the IPSat model with and without geometric fluctuations. We compare our results with the HERA measurements at $\langle W \rangle = 100 \gev$, corresponding to (in case of $J/\Psi$ photoproduction at $t=0$) $\xpom=9.6\cdot 10^{-4}$~\cite{Aktas:2005xu,Chekanov:2002xi,Chekanov:2002rm,Aktas:2003zi}, and $\langle W \rangle = 75 \gev$ that corresponds to slightly larger $\xpom=1.7\cdot 10^{-3}$~\cite{Alexa:2013xxa}.

\begin{figure}[htb]
\centering
		\includegraphics[width=0.5\textwidth]{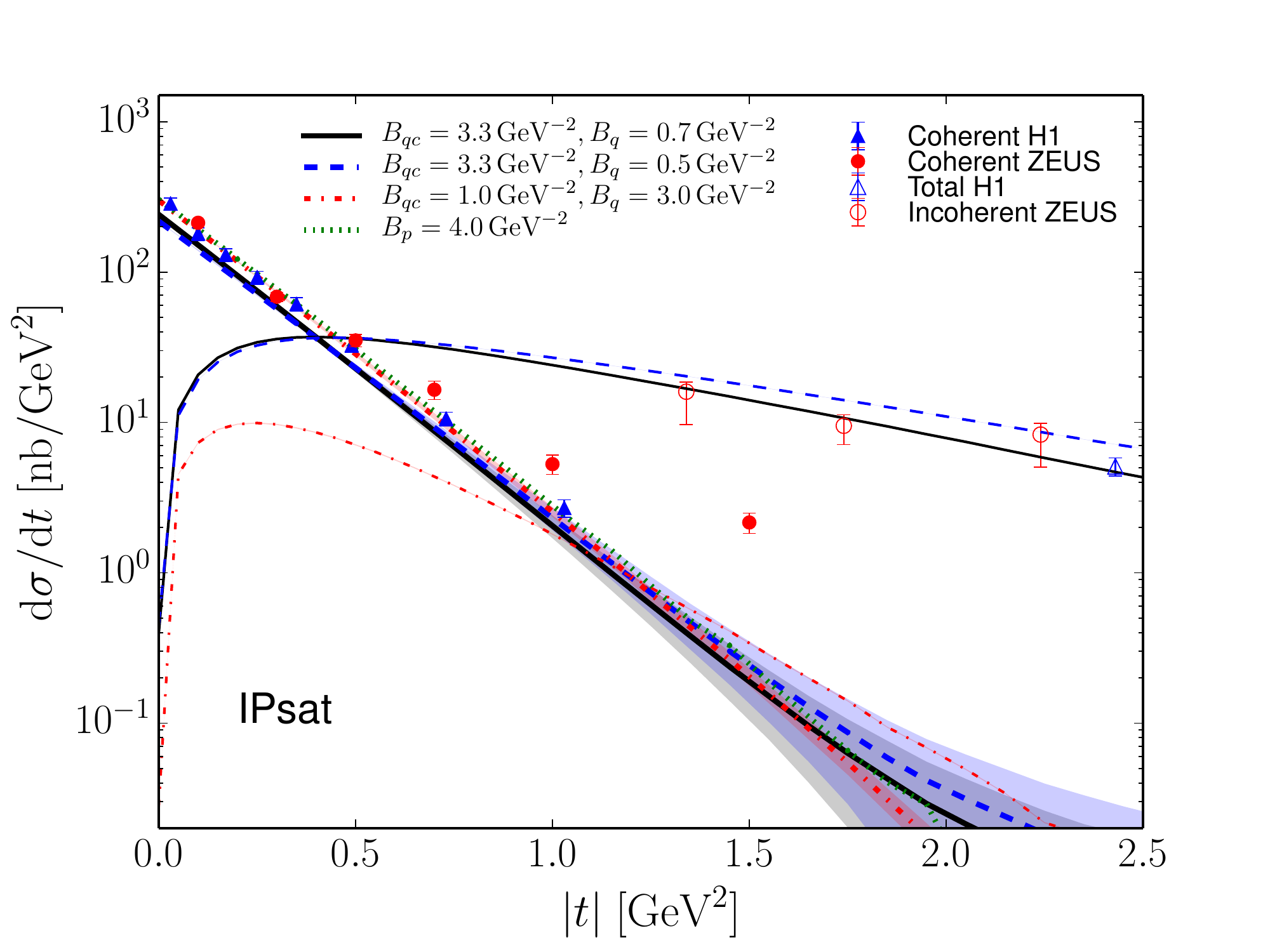} 
				\caption{Coherent (thick lines) and incoherent (thin lines) cross section as a function of $|t|$ compared with HERA data \cite{Aktas:2005xu,Chekanov:2002xi,Chekanov:2002rm,Aktas:2003zi}. The coherent cross section obtained without any fluctuations is also shown as a dotted line ($B_p=4.0\gev^{-2}$). The bands show statistical errors of the calculation.}
		\label{fig:bq-bp-10-35}
\end{figure}

Comparison to the H1 and ZEUS high energy data on coherent and incoherent diffractive $J/\Psi$ production as a function of $|t|$~\cite{Aktas:2005xu,Chekanov:2002xi,Chekanov:2002rm,Aktas:2003zi} at $\langle W \rangle = 100\gev$ is shown in Fig.\,\ref{fig:bq-bp-10-35}. At this energy, the H1 collaboration has measured the total diffractive cross section, which at high $|t|$ is to very good accuracy purely incoherent. 
Apart from the standard IPSat result with a round proton, for which the incoherent cross section is exactly zero, we employ the constituent quark profile discussed in Sec.\,\ref{sec:constituentquark}. We find that one has to introduce large geometric fluctuations (relatively small hot spots far away from the center of the proton with $B_{qc}=3.3\gev^{-2}, B_q=0.5 \dots 0.7\gev^{-2}$) in order to obtain a large enough variance and consequently a large incoherent cross section comparable with the experimental data. In particular, the much smoother proton configuration ($B_{qc}=1.0\gev^{-2}, B_q=3.0\gev^{-2}$) underestimates the incoherent cross section by several orders of magnitude while still being compatible with the measured coherent cross section. For typical proton configurations in these two situations see Fig.\,\ref{fig:ipsat_constituentquark_protons}.
One can further see that when the constituent quark size $B_q$ is decreased at constant $B_{qc}$ the amount of fluctuations increases leading to a larger incoherent cross section. Also, the $|t|$ slope of the incoherent cross section at large $|t|$ is directly given by the constituent quark size~\cite{Lappi:2010dd}.
Note that the overall normalization is affected by the inclusion of geometric fluctuations as discussed above.

\begin{figure}[tb]
\centering
		\includegraphics[width=0.5\textwidth]{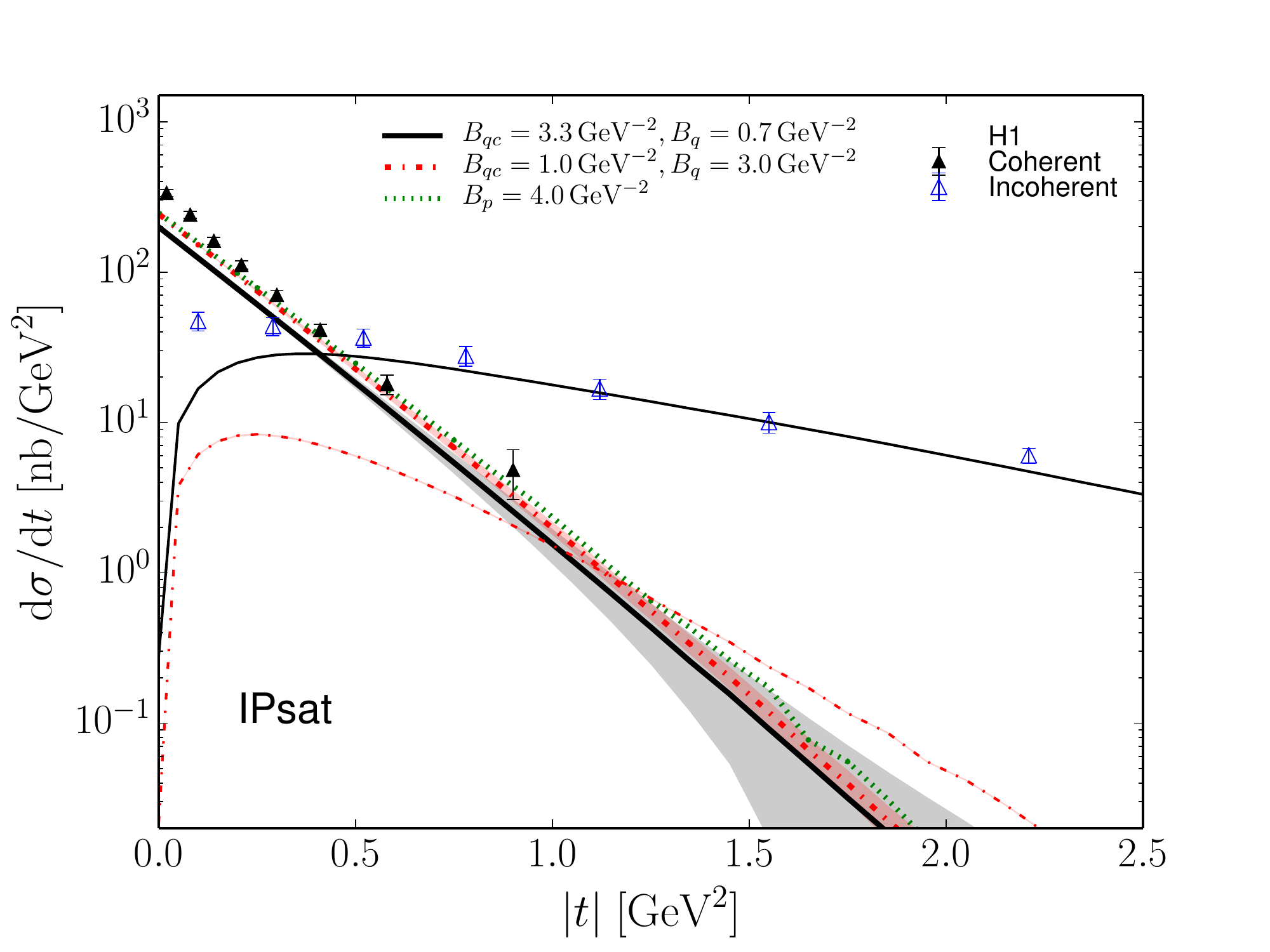} 
				\caption{Coherent (thick lines) and incoherent (thin lines) $J/\Psi$ production cross sections at $\langle W \rangle = 75 \gev$ compared with H1 data \cite{Alexa:2013xxa}. The result obtained without geometric fluctuations corresponds to $B_p=4.0\gev^{-2}$ line. }
		\label{fig:ipsat-w75-jpsi}
\end{figure}

Comparison to the H1 data~\cite{Alexa:2013xxa} at the lower $\langle W \rangle =75\gev$ is shown in Fig.\,\ref{fig:ipsat-w75-jpsi}. Conclusions are the same as for  $\langle W\rangle =100\gev$. The agreement with the lumpy proton structure ($B_{qc}=3.3\gev^{-2}, B_q=0.7 \gev^{-2}$), that also worked well with $\langle W \rangle = 100 \gev$ data, is good, while a smoother proton is incompatible with the incoherent data. 
We do not reproduce accurately the change in total coherent cross section from $\langle W \rangle=100 \gev$ to $\langle W \rangle=75 \gev$.
For the lumpy proton the incoherent cross section is only underestimated at very low $|t|$, where the contribution from e.g. saturation scale fluctuations is expected to be dominant~\cite{Miettinen:1978jb}. 
The effect of $Q_s$ fluctuations is studied numerically in Sec. \ref{sec:Qsfluc}.

\begin{figure}[tb]
\centering
		\includegraphics[width=0.5\textwidth]{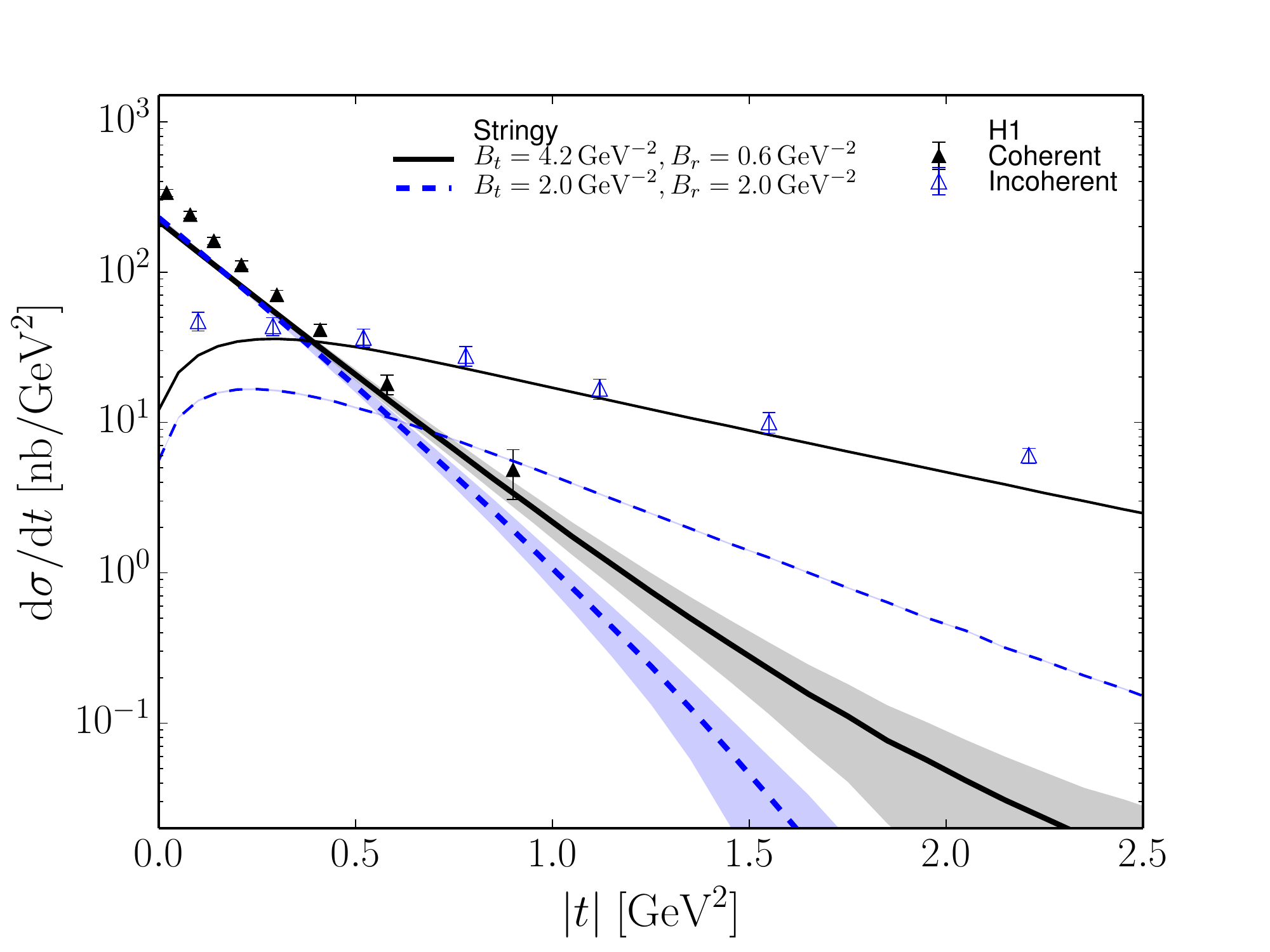} 
				\caption{Coherent (thick lines)  and incoherent (thin lines) cross section as a function of $|t|$ calculated using two ``stringy proton'' model parametrizations compared with H1 data \cite{Alexa:2013xxa}.  The bands show statistical errors of the calculation.}
		\label{fig:fluxtube-w-75}
\end{figure}

In order to study the dependence on the exact form of the geometric fluctuations, we next present diffractive cross sections calculated using the ``stringy proton'' density profile introduced in Sec. \ref{sec:fluxtube}. The results are shown in Fig.\,\ref{fig:fluxtube-w-75} and compared with H1 data at $\langle W \rangle = 75\gev$~\cite{Alexa:2013xxa} where we again see that we need large geometric fluctuations, 
corresponding to  a ``tube width'' $B_r$ much smaller than the average distance of the quarks from the center set by $B_t$. A good description of the data 
is obtained with $B_t=4.2\gev^{-2}$ and $B_r=0.6\gev^{-2}$. Example density profiles from the parametrization that has large fluctuations are shown in Fig.\,\ref{fig:fluxtube}.
A smoother parametrization that has $B_t=B_r$ is comparable with the coherent cross section measurements but underestimates the incoherent cross section by more than an order of magnitude.
 Comparing to the results obtained using constituent quark protons shown in Fig.\,\ref{fig:ipsat-w75-jpsi}, we conclude that the precise nature of the fluctuating shape cannot be constrained by the incoherent diffractive $J/\Psi$ production.

\begin{figure}[tb]
\centering
		\includegraphics[width=0.5\textwidth]{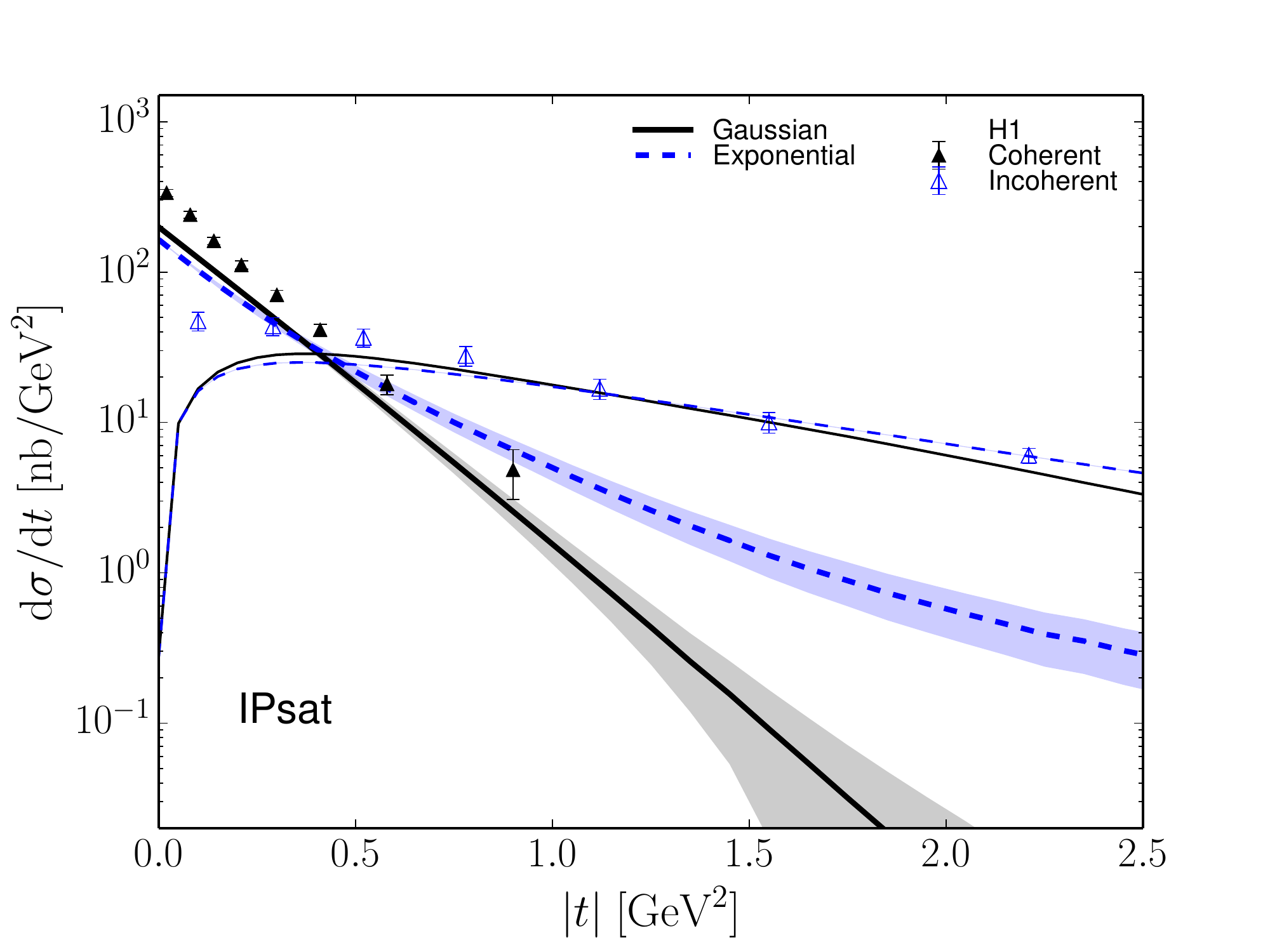} 
				\caption{Coherent (thick lines) and incoherent (thin lines) cross section as a function of $|t|$ calculated with Gaussian ($B_{qc}=3.3\gev^{-2}, B_q=0.7\gev^{-2}$) and exponential ($\tilde{B}_{qc}=0.91 \gev^{-1}, \tilde{B}_{q}=0.42\gev^{-1}$) density profile compared with HERA data~\cite{Alexa:2013xxa}.  The bands show statistical errors of the calculation.}
		\label{fig:exponential-w-100}
\end{figure}

The effect of replacing Gaussian density distributions by exponential distributions (see Eq.~\eqref{eq:exp}) in the constituent quark picture is shown in Fig.\,\ref{fig:exponential-w-100}. 
We obtain a good description of the H1 data with parameters $\tilde B_{qc}=0.91\gev^{-1}$ and $\tilde{B}_{q}=0.42\gev^{-1}$. With these parameters, we get the same 2-dimensional root mean square distance of the quark centers from the proton center as in case of the Gaussian distribution used in Fig.\,\ref{fig:exponential-w-100}. Similarly the quarks' 2-dimensional root-mean-square radii are the same.
This means that again we have large event-by-event fluctuations with 
 small constituent quarks far away from each other ($ \tilde B_{qc} \gg \tilde B_{q}$).
 Using exponential distributions mainly modifies the large $|t|$ tail of the coherent cross section. Because the coherent $|t|$ data ends at $|t|\sim 1\gev^2$, one cannot currently distinguish between Gaussian and exponential density profiles.

The coherent cross section is experimentally challenging to measure at large $|t|$ where the incoherent background starts to dominate. We can also see that in Fig.\,\ref{fig:bq-bp-10-35} the H1 and ZEUS results start to deviate in the largest $|t|$ bins. A precise measurement of the coherent cross section at large $|t|$ would allow us to further constrain the details of the average shape of the proton. Lacking such constraining data we choose to use a Gaussian distribution in the rest of this work.

\subsection{Including saturation scale fluctuations} \label{sec:Qsfluc}

\begin{figure}[tb]
\centering
		\includegraphics[width=0.5\textwidth]{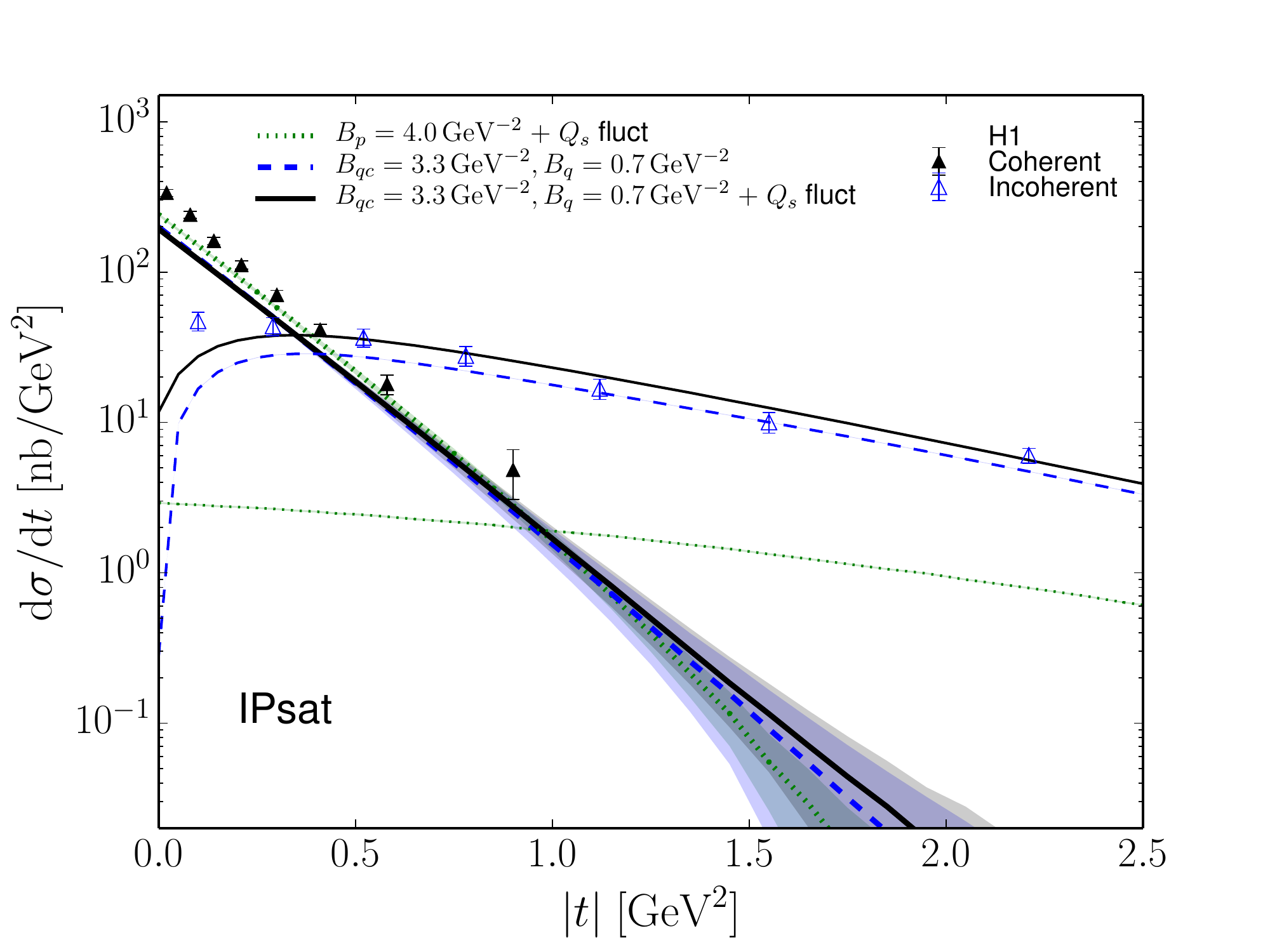} 
				\caption{Coherent (thick lines) and incoherent (thin lines) cross section as a function of $|t|$ compared with HERA data \cite{Alexa:2013xxa} at $\langle W \rangle = 75\gev$.  The bands show statistical errors of the calculation. Saturation scale fluctuations are included in the round proton case ($B_p=4.0\gev^{-2}$), and their effect on top of proton geometric fluctuations is also shown.}
\label{fig:bq-bp-10-35-qsfluct}
\end{figure}

Having analyzed the effect of geometric fluctuations we now turn to the study of additional fluctuations of the saturation scale. As described in Sec.\,\ref{sec:qsfluct} within the constituent quark proton model we allow the saturation scale of each quark to fluctuate individually. 
The spectra obtained with the same constituent quark proton parametrizations as used in Fig.\,\ref{fig:ipsat-w75-jpsi} and with additional saturation scale fluctuations are shown in Fig.\,\ref{fig:bq-bp-10-35-qsfluct}.
 In the figure we also show the cross sections obtained by allowing the saturation scale of a round proton ($B_p=4\gev^{-2}$) to fluctuate independently 
 between different cells of size $a^2=(0.4 \fm)^2$ in the transverse plane as discussed in Sec.\,\ref{sec:qsfluct}. 
As anticipated, we find that including saturation scale fluctuations improves the agreement with the experimental incoherent cross section, particularly at small $|t|$, with the effect diminishing at higher $|t|$. This is in line with early discussions of the effect of different kinds of fluctuations on incoherent diffraction \,\cite{Miettinen:1978jb}. The $Q_s$ fluctuations alone underestimate the measured incoherent cross section by approximately an order of magnitude.

In addition to $J/\Psi$, also diffractive production of lighter $\phi$ and $\rho$ mesons has been measured at HERA~\cite{Aid:1996ee,Adloff:1999kg,Breitweg:1999jy,Chekanov:2002rm,Chekanov:2005cqa,Aaron:2009xp}. The small mass of these mesons makes the photoproduction cross section calculation unreliable, because the cross section would receive significant contributions from large dipoles where non-perturbative effects become more relevant. 
The IPsat model includes some non-perturbative physics by requiring the dipole amplitude to reach unity in the large dipole limit. However, the model is still expected to reach the limits of its applicability as the dipole becomes large.
Thus, in the following we study the diffractive production of $\rho$ mesons at values of $Q^2$ that are large enough to allow for the perturbative treatment of the scattering process. However, even at $Q^2$ up to $\sim 20\gev^2$ the relative contribution from large dipoles
is stronger than in $J/\Psi$ photoproduction~\cite{Kowalski:2006hc}, which means that non-perturbative physics may be more relevant.

\begin{figure}[tb]
 \begin{minipage}{\linewidth}
      \centering
		\includegraphics[width=0.95\textwidth]{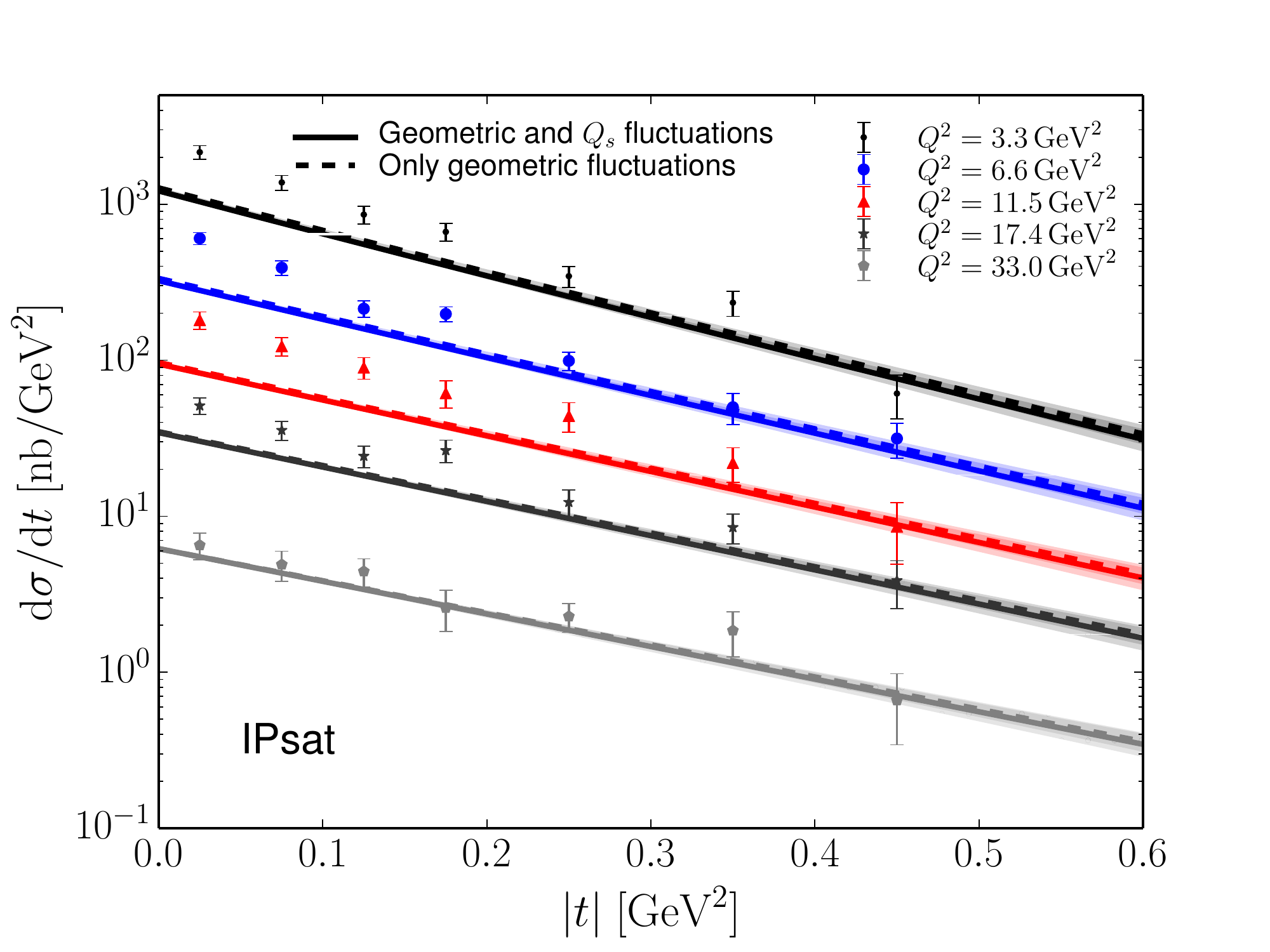} 
		\includegraphics[width=0.95\textwidth]{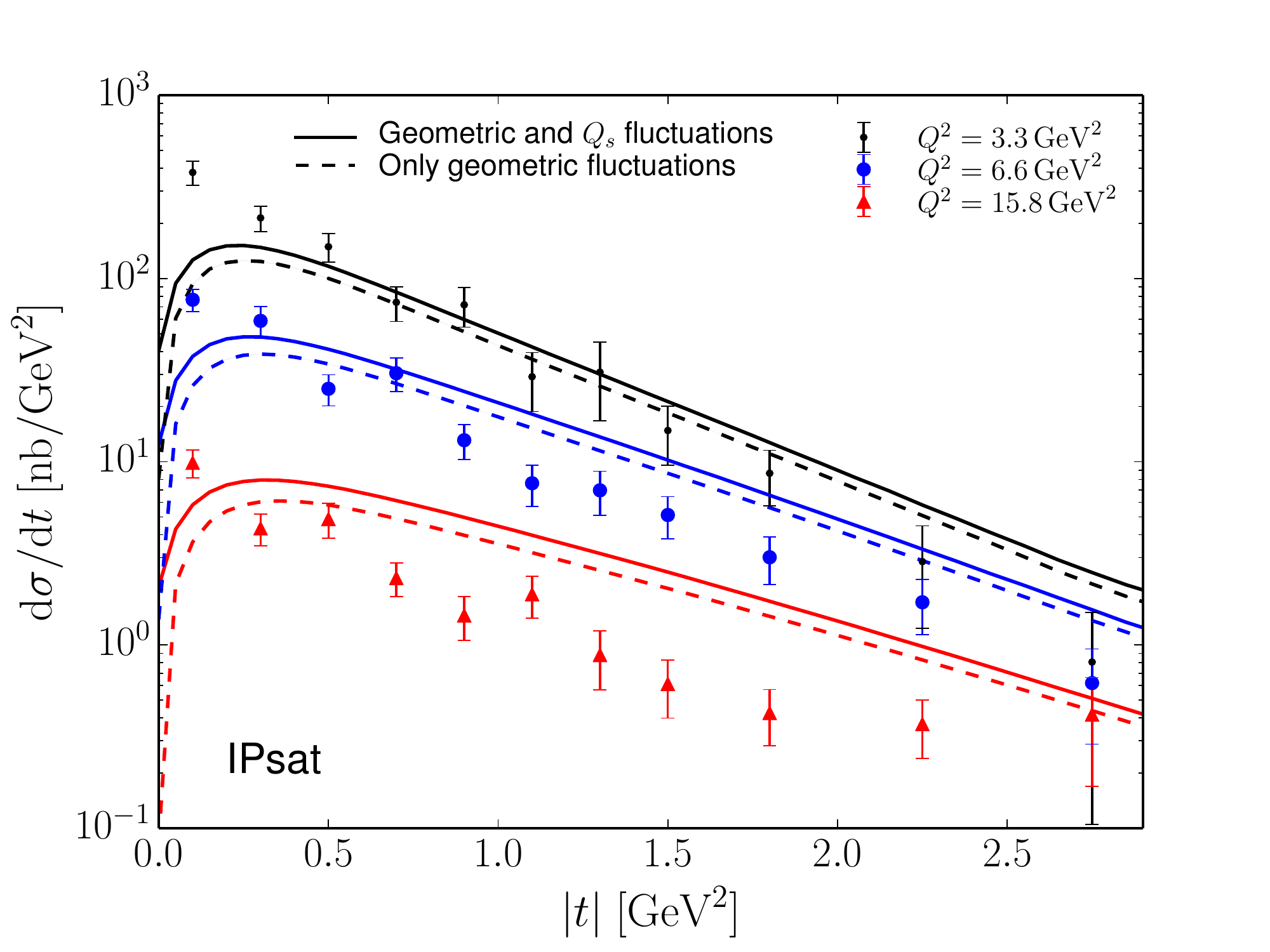} \label{fig:constituent_quark_rho_incoh}
\end{minipage}
\caption{ Coherent (upper) and incoherent (lower) diffractive $\rho$ production cross section at $W=75\gev$ as a function of $|t|$ compared with HERA data \cite{Aaron:2009xp}. The bands show statistical errors of the calculation. Geometric fluctuations are included using the constituent quark picture with $B_{qc}=3.3\gev^{-2}, B_q=0.7\gev^{-2}$. $Q_s$ fluctuations are included in the results represented by solid lines.}\label{fig:constituent_quark_rho}
\end{figure}

The H1 collaboration has measured coherent and incoherent $\rho$ production in the range $Q^2=3.3 \dots 33.0\gev^2$~\cite{Aaron:2009xp}. We calculate the corresponding cross sections within our framework by using the IPsat model with constituent quarks and the same parameters that were used to describe the $J/\Psi$ photoproduction data. The results are shown in Fig.\,\ref{fig:constituent_quark_rho} (upper panel) for coherent and in Fig.\,\ref{fig:constituent_quark_rho} (lower panel) for incoherent $\rho$ production.
For coherent cross section, the agreement with the data is better for the highest $Q^2$ bins.
For small $|t|$ the coherent cross section is underestimated, especially at low $Q^2$. The measured incoherent cross section would prefer slightly larger constituent quark size which would make the calculated $|t|$ slope steeper, but such a change would not be favored by the incoherent $J/\Psi$ production cross section which is theoretically under better control. As discussed above we expect our model to be less reliable in diffractive $\rho$ production due to contributions from large dipoles even at moderate values of $Q^2$.
When saturation scale fluctuations are included, the description of the small-$|t|$ part of the incoherent cross section is improved.

\subsection{IP-Glasma model}\label{sec:ipglasma}

\begin{figure}[tb]
\centering
		\includegraphics[width=0.5\textwidth]{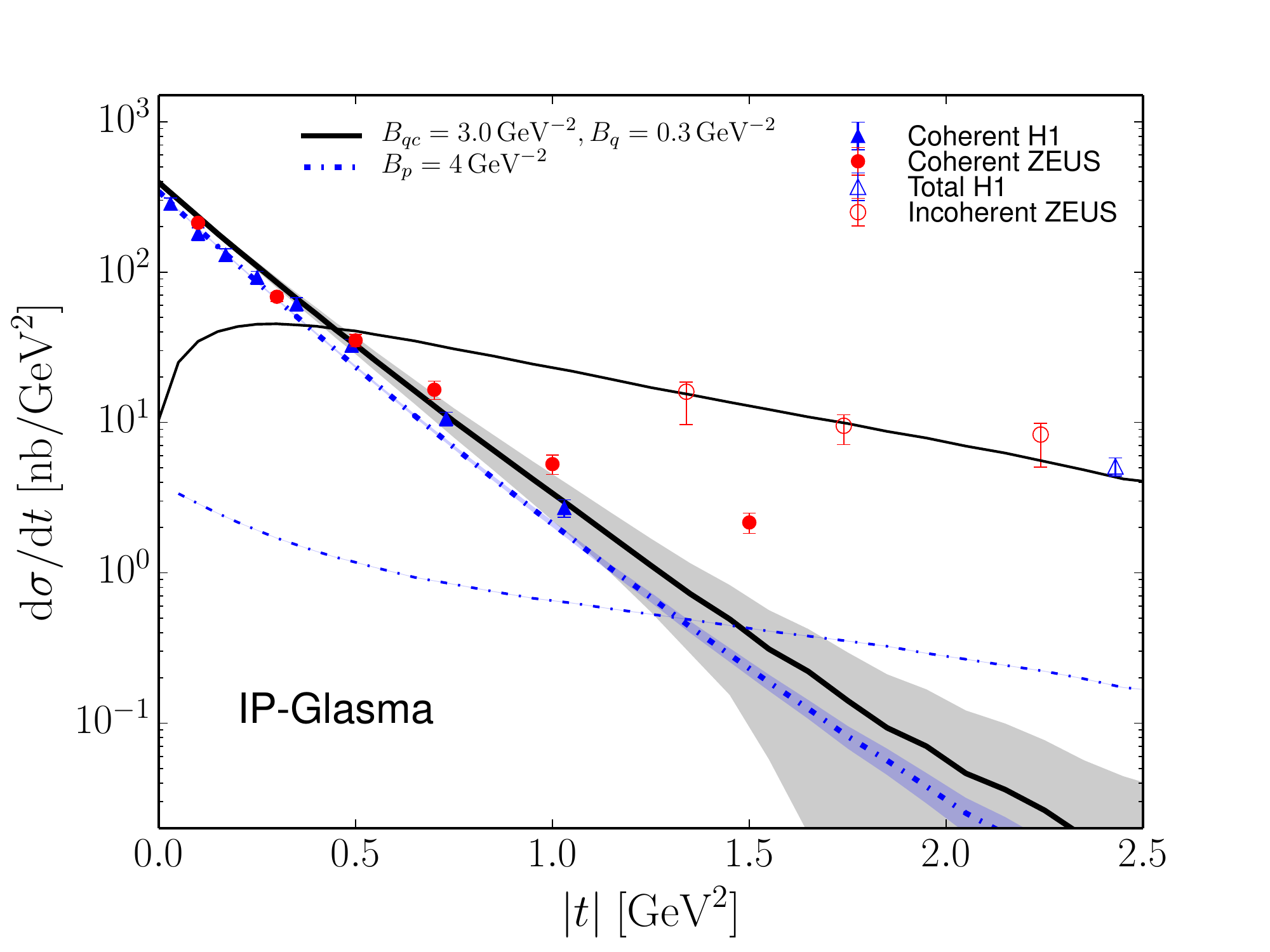} 
				\caption{Coherent (thick lines) and incoherent (thin lines) $J/\Psi$ photoproduction cross section in the IP-Glasma framework as a function of $|t|$ compared with HERA data \cite{Aktas:2005xu,Chekanov:2002xi,Chekanov:2002rm,Aktas:2003zi} at $\langle W\rangle=100 \gev$. 
			        The $B_p=4\gev^{-2}$ result includes only color charge fluctuations.}
		\label{fig:ipglasma_w_100}
\end{figure}

Finally, we present results for coherent and incoherent diffractive $J/\Psi$ and $\rho$ production in the IP-Glasma model.
The two main differences to the IPsat model are the existence of color charge fluctuations (in addition to possible saturation scale and geometric fluctuations), and the emergence of long-distance Coulomb tails in the gluon fields from the solution of the Yang-Mills equation. These infrared tails are regulated by the mass parameter $m$ (see Eq.\,(\ref{eq:wilson})) for which we use $m=0.4 \gev$. Other values of $m$ reduce the simultaneous agreement with experimental coherent and incoherent HERA data using any combination of parameters $B_{qc}$ and $B_q$ as we will demonstrate in Appendix \ref{app:m}. Other than these differences, and the fact that the dipole amplitude is computed from the Wilson lines (\ref{eq:wilson}) according to Eq.\,(\ref{eq:dipoleipglasma}) instead of Eq.\,(\ref{eq:unfactbt}), the physics content of the two models is the same. In particular, the geometry characterized by the different thickness functions is the same, only modified by the effects of large infrared tails in the IP-Glasma model, which are mainly compensated by the cutoff $m$.

First we compare coherent and incoherent cross sections with the HERA data for diffractive $J/\Psi$ production at $\langle W \rangle = 100 \gev$. The results are shown in Fig. \ref{fig:ipglasma_w_100}. We find that the color charge fluctuations alone are not enough to describe the large incoherent cross section. Large geometric fluctuations ($B_{qc} \gg  B_q$) on top of color charge fluctuations are needed to obtain an incoherent cross section compatible with the experimental data\footnote{Note that in our previous calculation in \cite{Mantysaari:2016ykx} the center of the proton was moved to the origin after the constituent quark positions were sampled, effectively making the proton smaller. This transformation is not done in this work which changes the numerical value of $B_{qc}$. Concerning the related issue of retaining the proton's center of mass see \cite{Mitchell:2016jio}.
}. 
Because $m$ does affect the size of the system, it is the combination of $B_{qc}$, $B_q$ and $m$ that determines the geometry and its fluctuations in the IP-Glasma model. A direct comparison of $B_{qc}$ and $B_q$ between IPSat and IP-Glasma is thus difficult.

\begin{figure}[tb]
\centering
		\includegraphics[width=0.5\textwidth]{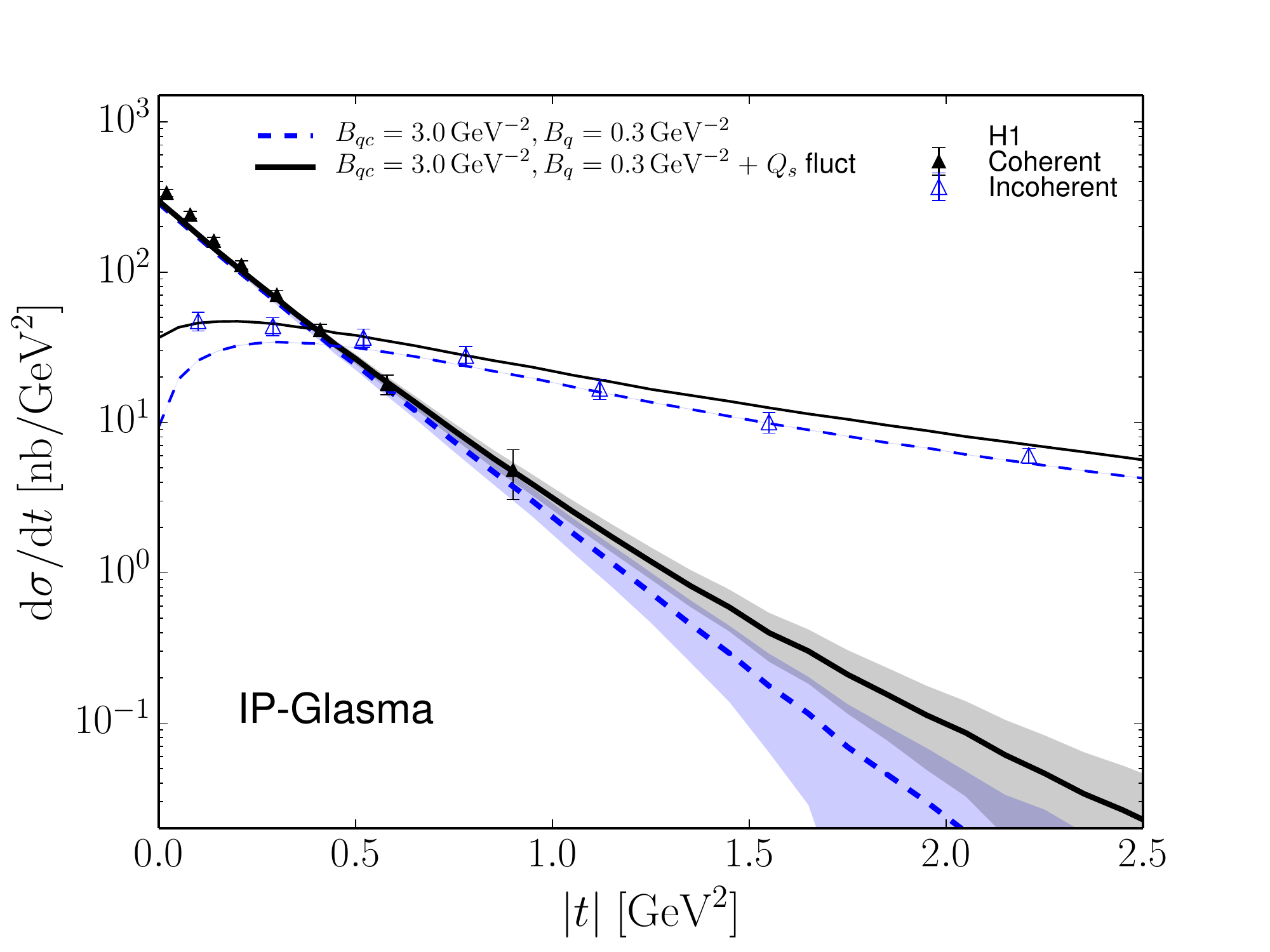} 
				\caption{Coherent (thick lines) and incoherent (thin lines) $J/\Psi$ production cross sections at $\langle W \rangle = 75 \gev$ compared with H1 data \cite{Alexa:2013xxa}. }
		\label{fig:ipglasma_jpsi_w_75}
\end{figure}

We next compare with H1 data at $\langle W \rangle = 75 \gev$, where the incoherent cross section is measured also at smaller $|t|$~\cite{Alexa:2013xxa}. The results are shown in Fig.\,\ref{fig:ipglasma_jpsi_w_75}. We find again that only when including large geometric fluctuations is the large-$|t|$ part of the incoherent cross section described well. The small-$|t|$ part of the incoherent cross section can only be reproduced with additional saturation scale fluctuations. This was expected based on Ref.\,\cite{Miettinen:1978jb}, as saturation scale fluctuations contribute to the incoherent cross section dominantly at small $|t|$. 
Fluctuations at different distance scales are visible in the incoherent cross section: the lowest-$|t|$ part is sensitive to $Q_s$ fluctuations that are visible at the largest distance scales, as they correspond to fluctuations of overall density. Geometrical fluctuations become dominant at $|t|\gtrsim 0.2 \gev^2$, where we become sensitive to distance scales smaller than the proton size. 
Color charge fluctuations take place at very small distance scales but also affect the overall normalization. As shown explicitly in Fig.\,\ref{fig:ipglasma_w_100}, their effect is thus mainly visible at very small $|t|$. 
Overall, including geometric, $Q_s$, and color charge fluctuations in the IP-Glasma model, we are able to achieve excellent agreement with the experimental data at all values of $|t|$.

\begin{figure}
 \begin{minipage}{\linewidth}
      \centering
		\includegraphics[width=0.95\textwidth]{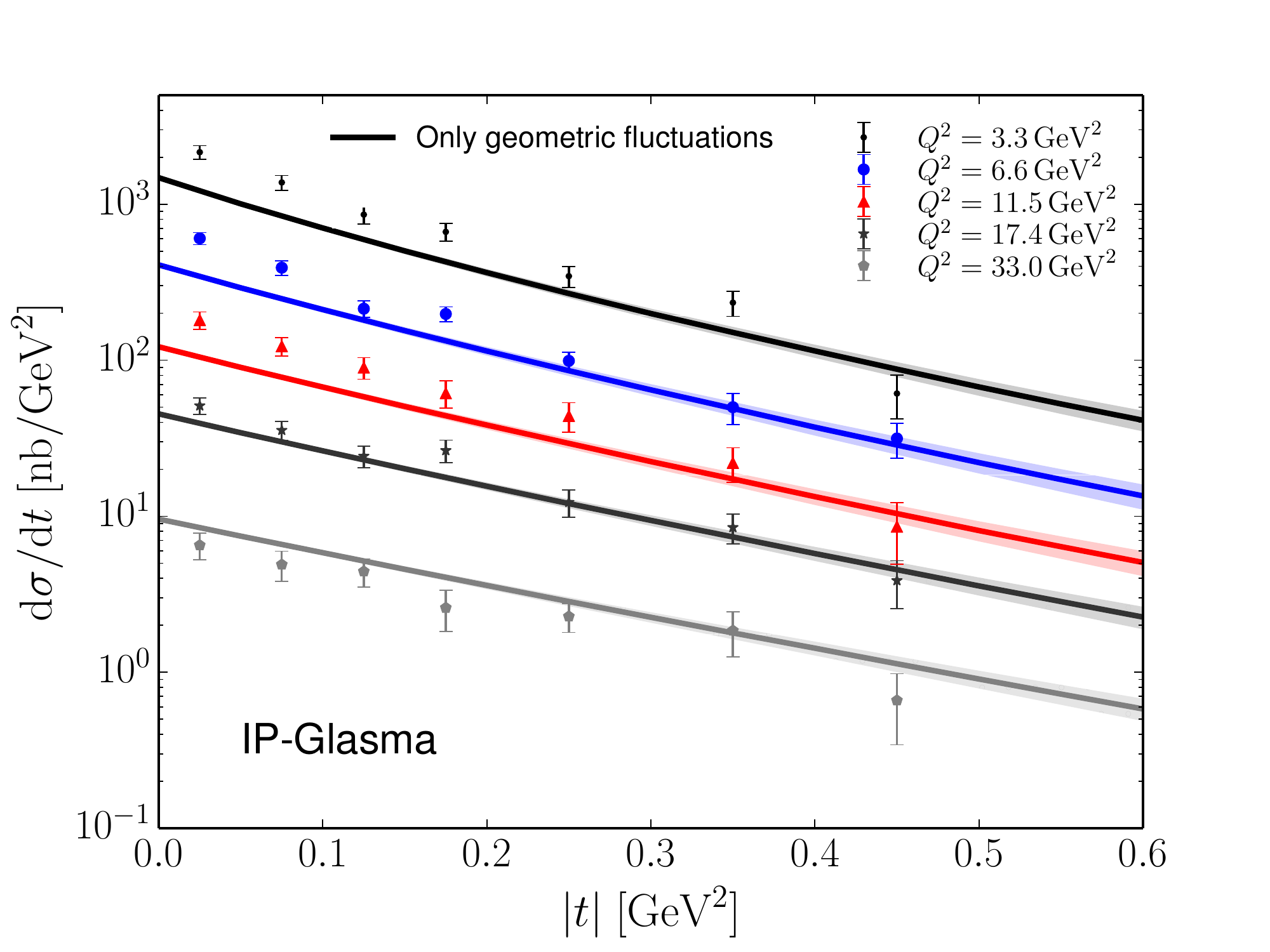} 
		\includegraphics[width=0.95\textwidth]{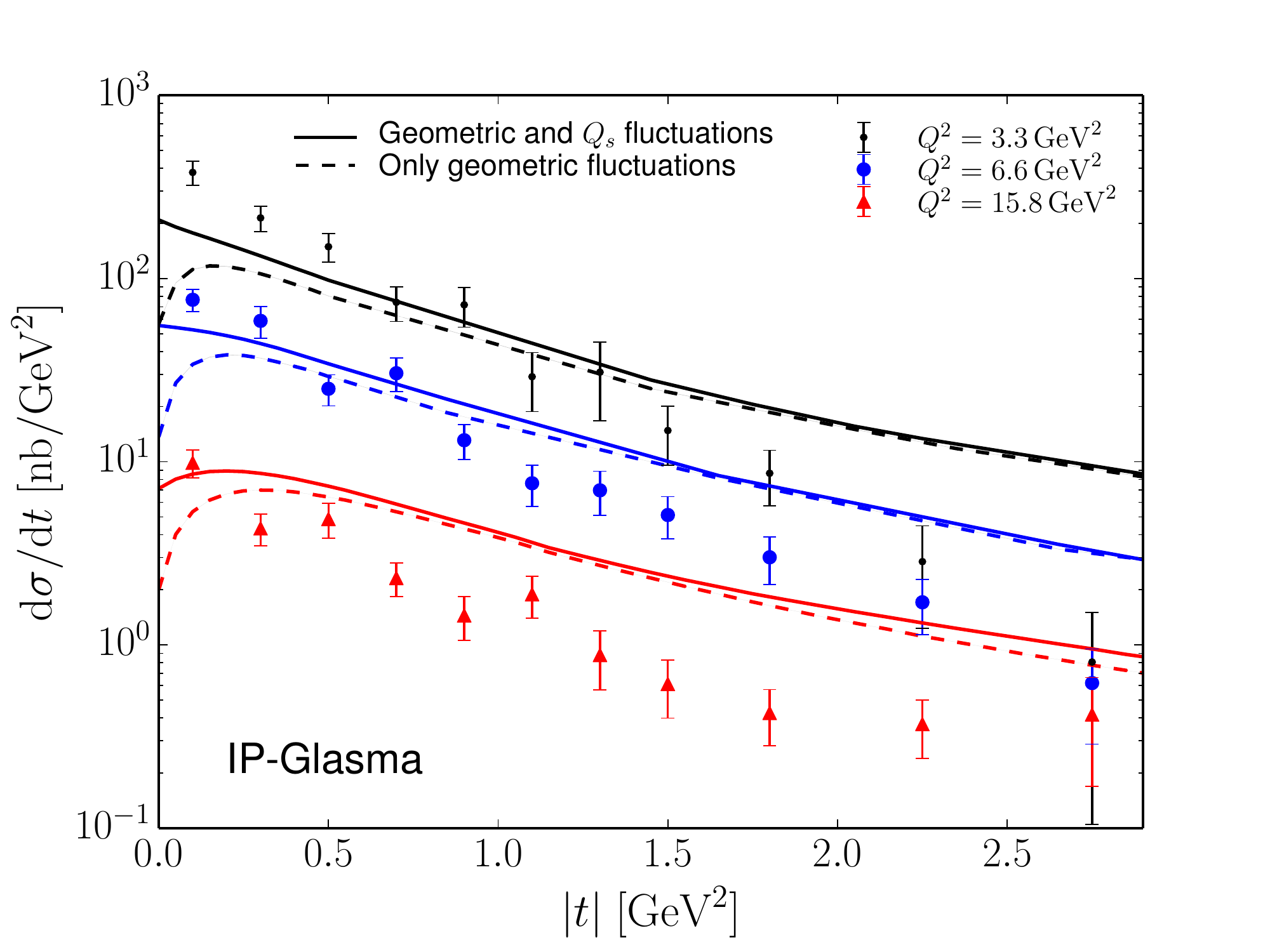}
		\end{minipage}
	\caption{Coherent (upper) and incoherent (lower) diffractive $\rho$ production cross section at $\langle W\rangle =75\gev$ as a function of $|t|$ compared with HERA data~\cite{Aaron:2009xp} calculated in the IP-Glasma framework. The bands show statistical errors of the calculation. Geometric fluctuations are included using the constituent quark picture with $B_{qc}=3.0\gev^{-2}, B_q=0.3\gev^{-2}$ and $m=0.4\gev$. For the incoherent cross section the effect of $Q_s$ fluctuations is included in the result shown as solid curves. }
	\label{fig:ipglasma-rho}
\end{figure}

The $\rho$ production cross sections calculated using the same fluctuating proton parametrizations are shown in Fig.\,\ref{fig:ipglasma-rho}. The coherent cross section measured at large $Q^2$ is described well, and $Q_s$ fluctuations are again found to improve the description of incoherent cross section data at small $|t|$. Neither the coherent cross section at small $Q^2$ nor the incoherent cross section at large $|t|$ are described accurately by our calculation. This is likely due to contributions from large dipoles that are not correctly described within our framework as discussed earlier in case of the IPsat model. In the IP-Glasma model the situation is even worse as the dipole cross section does not go to one at large $r$ like in the parametrized IPSat expression.  This is evident from Eq.\,(\ref{eq:dipoleipglasma}): as soon as one end of the dipole is outside the proton, the expression for $N$ goes to zero (also see Ref.~\cite{Schlichting:2014ipa}).

\section{Conclusions and Outlook}
We have presented a detailed event-by-event computation of exclusive diffractive vector meson production in the color glass condensate framework.
Within the IPSat and IP-Glasma models, whose parameters are almost entirely constrained by HERA data on deeply inelastic scattering, we find that in order to describe the experimental incoherent cross section of both $J/\Psi$ and $\rho$ production, large geometric fluctuations are needed. This finding is independent of the details of the model. These include different density distributions of gluons in the proton, of which we studied Gaussian and exponential distributions, as well as a stringy model, motivated by QCD in the limit of large quark masses.
Apart from geometric fluctuations, we included fluctuations of the saturation scale and in the case of the IP-Glasma model, color charges. They contribute at all values of $|t|$ but dominate in the limit $|t|\rightarrow 0$.
In particular in the IP-Glasma model, which includes all relevant fluctuations, we find excellent agreement of both the coherent and incoherent diffractive $J/\Psi$ production cross sections. The diffractive production of $\rho$ mesons is less accurately described, which we can attribute to more significant contributions from large dipoles that are not  well described in our framework.

Our analysis provides constraints on the proton's fluctuating shape at high energy (small $x$), which is an important input for calculations of observables in p+p and p+A collisions. These include in particular azimuthal anisotropy coefficients, which in case of strong final state effects are highly sensitive to the initial shape of the proton. We will investigate in the future if the fluctuating proton shape constrained in this work is indeed compatible with experimental data on anisotropic flow in p+Pb collisions at the LHC  and p+Au collisions at RHIC.

\section*{Acknowledgments}
We thank E. Aschenauer, T. Lappi, S. Schlichting, M. Strikman, T. Ullrich,  and R. Venugopalan for discussions.
This work was supported under DOE Contract No. DE-SC0012704. This research used resources of the National Energy Research Scientific Computing Center, which is supported by the Office of Science of the U.S. Department of Energy under Contract No. DE-AC02-05CH11231. BPS acknowledges a DOE Office of Science Early Career Award.

\appendix

\section{Phenomenological corrections}\label{appendix:corrections}

\begin{figure}[tb]
\centering
		\includegraphics[width=0.5\textwidth]{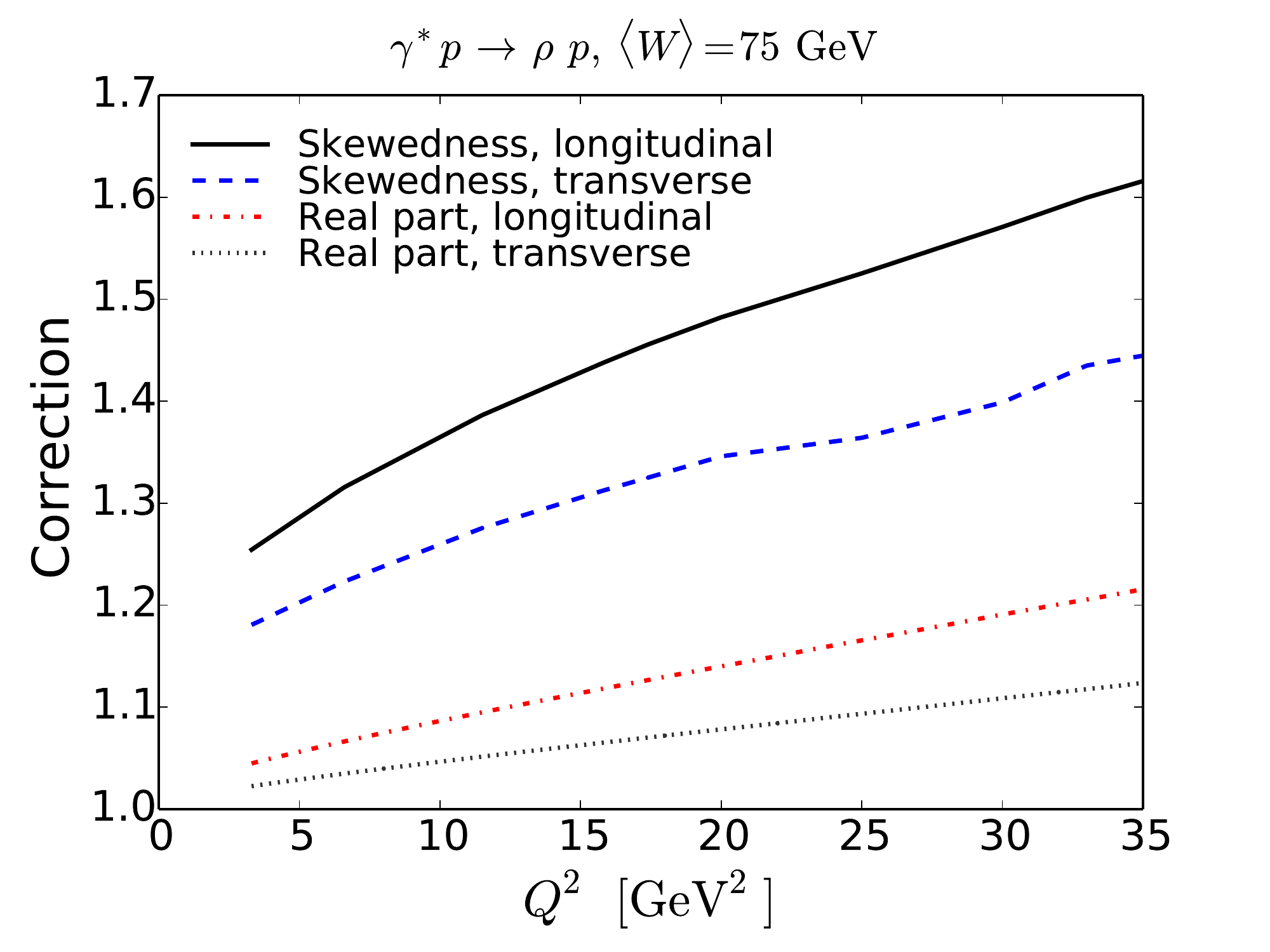} 
				\caption{Average effect of the skewedness and real part corrections at $|t|<0.5 \gev^2$ to the coherent $\rho$ production cross section calculated from the IPsat model without fluctuations at $\langle W\rangle =75\gev$. }
		\label{fig:skewedness_rho}
\end{figure}

As discussed already in Sec. \ref{sec:corrections}, the phenomenological corrections, and especially the skewedness correction, are numerically important. To demonstrate this, we show in Fig.\,\ref{fig:skewedness_rho} the effect of the skewedness and real part corrections on the coherent diffractive $\rho$ production cross section at different values of $Q^2$. The corrections are calculated separately for transversally and longitudinally polarized photons. The effect of the skewedness correction (see Eq.~\eqref{eq:skew}) is quantified in the IPSat model without fluctuations by the ratio of the diffractive $\rho$ production cross section with and without taking the skewedness correction into account. The real part correction is quantified by the factor $(1+\beta^2)$ from Eq.~\eqref{eq:realpart_beta}, again calculated in the IPSat model without fluctuations.

As the corrections depend slightly on $|t|$, the results shown in Fig.\,\ref{fig:skewedness_rho} are the average correction factors at $|t|<0.5 \gev^2$. We observe that especially at high $Q^2$, where the gluon density rises most rapidly, the skewedness correction becomes very large, of the order of $50\%$. For $J/\Psi$ photoproduction in the same kinematics (as shown in Fig.\,\ref{fig:bq-bp-10-35-qsfluct}) the skewedness correction is $\approx 43\%$ and the real part correction $\approx 11 \%$. 

\section{Dependence on the number of constituent quarks}\label{appendix:nq}
\begin{figure}[tb]
\centering
		\includegraphics[width=0.5\textwidth]{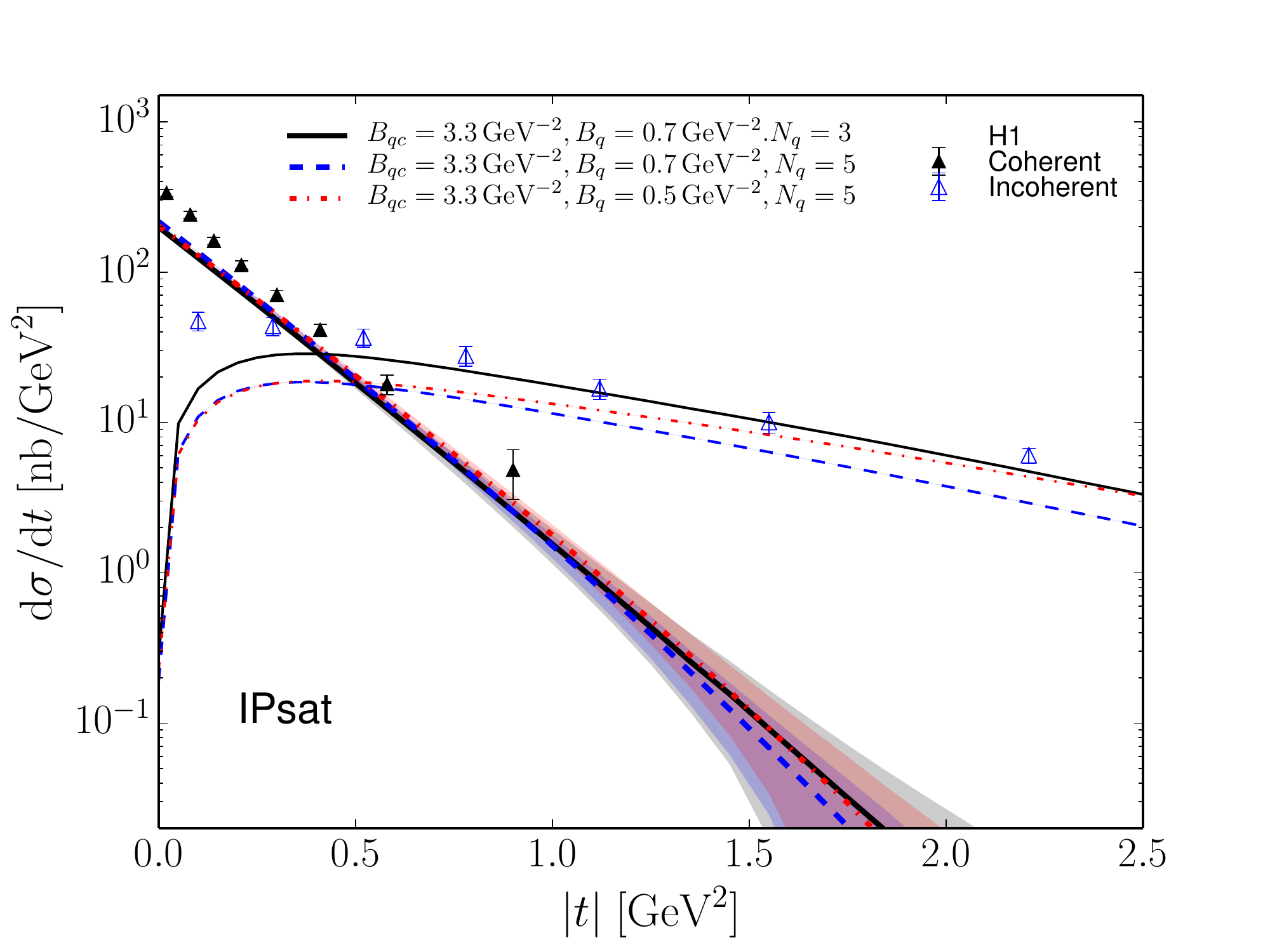} 
				\caption{Dependence of coherent (thick lines) and incoherent (thin lines) diffractive $J/\Psi$ production cross section at $\langle W\rangle=75\gev$ on the number of constituent quarks (hot spots) $N_q$.    The bands show statistical errors of the calculation. }
		\label{fig:nq_dep}
\end{figure}

We study the dependence of the effect of geometric fluctuations on the number of hot spots ($N_q$ in Eq.~\eqref{eq:TqReplace}). Numbers larger than 3 can be interpreted as the three constituent quarks plus large $x$ sea-quarks or gluons, which are emitted from the large-$x$ valence quarks (see also Ref.~\cite{Flensburg:2012zy,Mueller:2014fba}). This change does not affect the coherent cross section, as the average proton density profile remains approximately the same. 
However, it results in a smoother proton on average and thus one would expect to see a smaller incoherent cross section compared to the case with $N_q=3$. This is demonstrated in Fig.\,\ref{fig:nq_dep}, where having $N_q=5$ hot spots decreases the incoherent cross section by $\sim 30\%$. If the size of the hot spots is reduced by decreasing the constituent quark width from $B_q=0.7 \gev^{-2}$ to $B_q=0.5\gev^{-2}$, a similar degree of fluctuations and comparable incoherent cross section is obtained at large $|t|$. $Q_s$ fluctuations are not included in this analysis.

\section{Dependence on the infrared cutoff in the IP-Glasma model}\label{app:m}

\begin{figure}[tb]
\centering
		\includegraphics[width=0.5\textwidth]{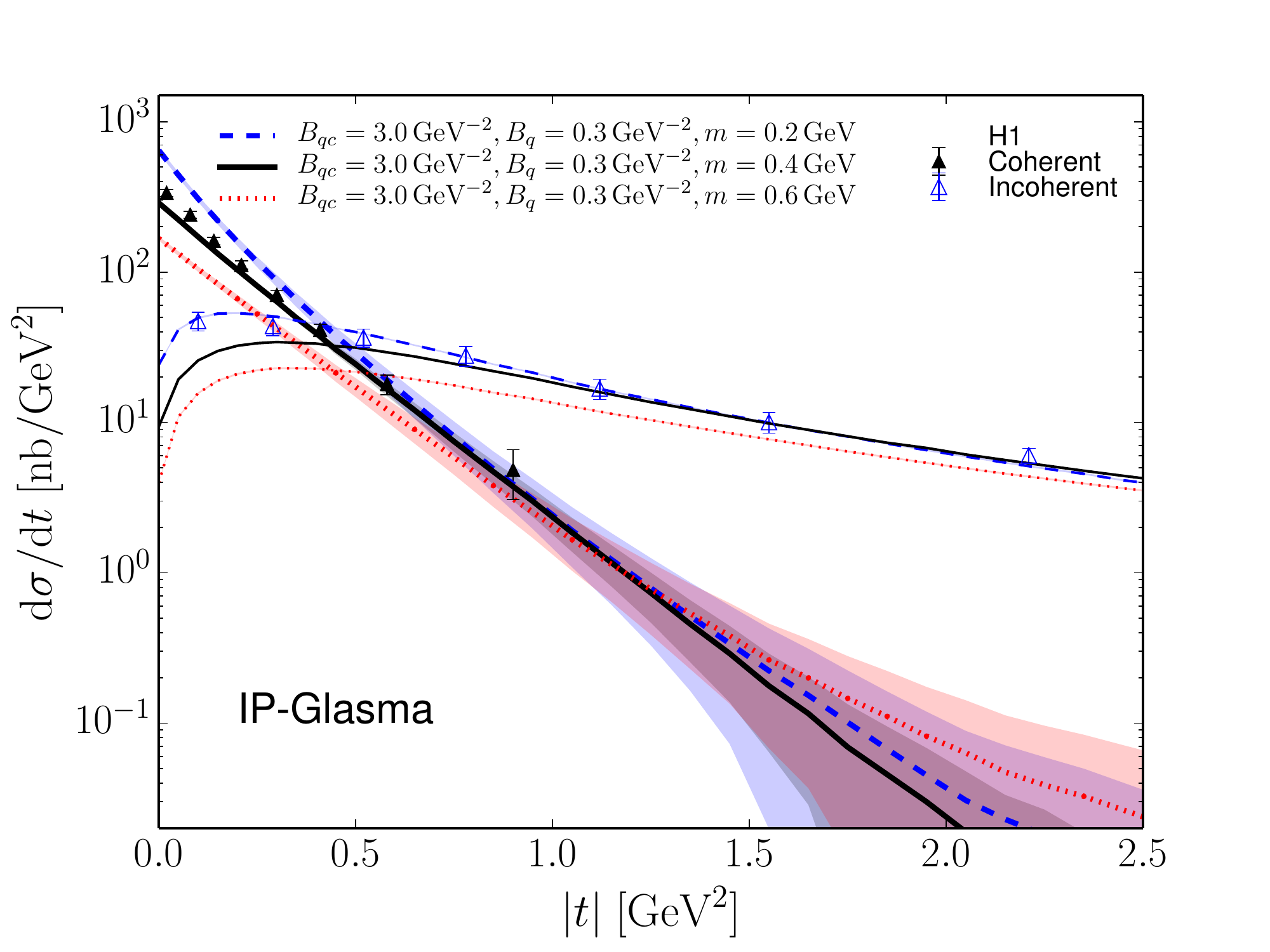} 
				\caption{Sensitivity of the results on the infrared cutoff $m$ in the IP-Glasma model. Cross sections are compared with HERA data \cite{Alexa:2013xxa}. }
		\label{fig:ipglasma_mdep}
\end{figure}

Because it affects the average proton size and overall normalization of the gluon distribution, the infrared cutoff parameter $m$, introduced for the IP-Glasma model in Eq.~\eqref{eq:wilson}, is expected to have an effect on both the coherent and incoherent cross sections. To study the sensitivity on this parameter, we show in Fig.\,\ref{fig:ipglasma_mdep} the coherent and incoherent cross section calculated with $m$ ranging from $0.2\,\gev$ to $0.6\,\gev$. 
As could be expected, the results are most sensitive to the infrared cutoff in the small-$|t|$ region, while for $|t|\gtrsim 1 \gev^2$ its effect becomes negligible. The dependence on $m$ at small momentum can be understood as follows: Smaller masses allow for longer Coulomb tails, making the proton effectively larger, leading to steeper coherent $|t|$ spectra. The fact that for smaller $m$ the proton becomes more dense at large impact parameters also increases the overall normalization of both the coherent and incoherent cross sections. We note that the ratio of the incoherent and the coherent cross section is almost independent of $m$ \cite{Mantysaari:2016ykx}.

\bibliographystyle{JHEP-2modlong.bst}
\bibliography{./refs}

\providecommand{\href}[2]{#2}\begingroup\raggedright\begin{thebibliography}{100}

\bibitem{Aaron:2009aa}
{\bf H1 and ZEUS} collaboration, F.~Aaron {\em et.~al.}, {\it {Combined
  Measurement and QCD Analysis of the Inclusive $e^\pm p$ Scattering Cross
  Sections at HERA}},  \href{http://dx.doi.org/10.1007/JHEP01(2010)109}{{\em
  JHEP} {\bf 1001} (2010) 109} [\href{http://arXiv.org/abs/0911.0884}{{\tt
  arXiv:0911.0884 [hep-ex]}}].

\bibitem{Abramowicz:2015mha}
{\bf H1 and ZEUS} collaboration, H.~Abramowicz {\em et.~al.}, {\it {Combination
  of measurements of inclusive deep inelastic ${e^{\pm }p}$ scattering cross
  sections and QCD analysis of HERA data}},
  \href{http://dx.doi.org/10.1140/epjc/s10052-015-3710-4}{{\em Eur. Phys. J.}
  {\bf C75} (2015) 580} [\href{http://arXiv.org/abs/1506.06042}{{\tt
  arXiv:1506.06042 [hep-ex]}}].

\bibitem{Belitsky:2003nz}
A.~V. Belitsky, X.-d. Ji and F.~Yuan, {\it {Quark imaging in the proton via
  quantum phase space distributions}},
  \href{http://dx.doi.org/10.1103/PhysRevD.69.074014}{{\em Phys. Rev.} {\bf
  D69} (2004) 074014} [\href{http://arXiv.org/abs/hep-ph/0307383}{{\tt
  arXiv:hep-ph/0307383 [hep-ph]}}].

\bibitem{Mueller:1998fv}
D.~Müller, D.~Robaschik, B.~Geyer, F.~M. Dittes and J.~Hořejši, {\it {Wave
  functions, evolution equations and evolution kernels from light ray operators
  of QCD}},  \href{http://dx.doi.org/10.1002/prop.2190420202}{{\em Fortsch.
  Phys.} {\bf 42} (1994) 101} [\href{http://arXiv.org/abs/hep-ph/9812448}{{\tt
  arXiv:hep-ph/9812448 [hep-ph]}}].

\bibitem{Ji:1996nm}
X.-D. Ji, {\it {Deeply virtual Compton scattering}},
  \href{http://dx.doi.org/10.1103/PhysRevD.55.7114}{{\em Phys. Rev.} {\bf D55}
  (1997) 7114} [\href{http://arXiv.org/abs/hep-ph/9609381}{{\tt
  arXiv:hep-ph/9609381 [hep-ph]}}].

\bibitem{Radyushkin:1997ki}
A.~V. Radyushkin, {\it {Nonforward parton distributions}},
  \href{http://dx.doi.org/10.1103/PhysRevD.56.5524}{{\em Phys. Rev.} {\bf D56}
  (1997) 5524} [\href{http://arXiv.org/abs/hep-ph/9704207}{{\tt
  arXiv:hep-ph/9704207 [hep-ph]}}].

\bibitem{Collins:1996fb}
J.~C. Collins, L.~Frankfurt and M.~Strikman, {\it {Factorization for hard
  exclusive electroproduction of mesons in QCD}},
  \href{http://dx.doi.org/10.1103/PhysRevD.56.2982}{{\em Phys. Rev.} {\bf D56}
  (1997) 2982} [\href{http://arXiv.org/abs/hep-ph/9611433}{{\tt
  arXiv:hep-ph/9611433 [hep-ph]}}].

\bibitem{Tangerman:1994eh}
R.~D. Tangerman and P.~J. Mulders, {\it {Intrinsic transverse momentum and the
  polarized Drell-Yan process}},
  \href{http://dx.doi.org/10.1103/PhysRevD.51.3357}{{\em Phys. Rev.} {\bf D51}
  (1995) 3357} [\href{http://arXiv.org/abs/hep-ph/9403227}{{\tt
  arXiv:hep-ph/9403227 [hep-ph]}}].

\bibitem{Mulders:2000sh}
P.~J. Mulders and J.~Rodrigues, {\it {Transverse momentum dependence in gluon
  distribution and fragmentation functions}},
  \href{http://dx.doi.org/10.1103/PhysRevD.63.094021}{{\em Phys. Rev.} {\bf
  D63} (2001) 094021} [\href{http://arXiv.org/abs/hep-ph/0009343}{{\tt
  arXiv:hep-ph/0009343 [hep-ph]}}].

\bibitem{Kimber:2001sc}
M.~A. Kimber, A.~D. Martin and M.~G. Ryskin, {\it {Unintegrated parton
  distributions}},  \href{http://dx.doi.org/10.1103/PhysRevD.63.114027}{{\em
  Phys. Rev.} {\bf D63} (2001) 114027}
  [\href{http://arXiv.org/abs/hep-ph/0101348}{{\tt arXiv:hep-ph/0101348
  [hep-ph]}}].

\bibitem{D'Alesio:2004up}
U.~D'Alesio and F.~Murgia, {\it {Parton intrinsic motion in inclusive particle
  production: Unpolarized cross sections, single spin asymmetries and the
  Sivers effect}},  \href{http://dx.doi.org/10.1103/PhysRevD.70.074009}{{\em
  Phys. Rev.} {\bf D70} (2004) 074009}
  [\href{http://arXiv.org/abs/hep-ph/0408092}{{\tt arXiv:hep-ph/0408092
  [hep-ph]}}].

\bibitem{Anselmino:2004ky}
M.~Anselmino, M.~Boglione, U.~D'Alesio, E.~Leader and F.~Murgia, {\it {Parton
  intrinsic motion: Suppression of the Collins mechanism for transverse single
  spin asymmetries in $p^\uparrow p \to \pi X$}},
  \href{http://dx.doi.org/10.1103/PhysRevD.71.014002}{{\em Phys. Rev.} {\bf
  D71} (2005) 014002} [\href{http://arXiv.org/abs/hep-ph/0408356}{{\tt
  arXiv:hep-ph/0408356 [hep-ph]}}].

\bibitem{Anselmino:2005sh}
M.~Anselmino, M.~Boglione, U.~D'Alesio, E.~Leader, S.~Melis and F.~Murgia, {\it
  {The general partonic structure for hadronic spin asymmetries}},
  \href{http://dx.doi.org/10.1103/PhysRevD.73.014020}{{\em Phys. Rev.} {\bf
  D73} (2006) 014020} [\href{http://arXiv.org/abs/hep-ph/0509035}{{\tt
  arXiv:hep-ph/0509035 [hep-ph]}}].

\bibitem{Aybat:2011zv}
S.~M. Aybat and T.~C. Rogers, {\it {TMD Parton Distribution and Fragmentation
  Functions with QCD Evolution}},
  \href{http://dx.doi.org/10.1103/PhysRevD.83.114042}{{\em Phys. Rev.} {\bf
  D83} (2011) 114042} [\href{http://arXiv.org/abs/1101.5057}{{\tt
  arXiv:1101.5057 [hep-ph]}}].

\bibitem{Mantysaari:2016ykx}
H.~Mäntysaari and B.~Schenke, {\it {Evidence of strong proton shape
  fluctuations from incoherent diffraction}},
  \href{http://arXiv.org/abs/1603.04349}{{\tt arXiv:1603.04349 [hep-ph]}}.

\bibitem{Iancu:2003xm}
E.~Iancu and R.~Venugopalan, {\it {The Color Glass Condensate and high-energy
  scattering in QCD}},  \href{http://arXiv.org/abs/hep-ph/0303204}{{\tt
  arXiv:hep-ph/0303204 [hep-ph]}}.

\bibitem{Gelis:2010nm}
F.~Gelis, E.~Iancu, J.~Jalilian-Marian and R.~Venugopalan, {\it {The Color
  Glass Condensate}},
  \href{http://dx.doi.org/10.1146/annurev.nucl.010909.083629}{{\em Ann. Rev.
  Nucl. Part. Sci.} {\bf 60} (2010) 463}
  [\href{http://arXiv.org/abs/1002.0333}{{\tt arXiv:1002.0333 [hep-ph]}}].

\bibitem{Albacete:2010sy}
J.~L. Albacete, N.~Armesto, J.~G. Milhano, P.~Quiroga-Arias and C.~A. Salgado,
  {\it {AAMQS: A non-linear QCD analysis of new HERA data at small-x including
  heavy quarks}},  \href{http://dx.doi.org/10.1140/epjc/s10052-011-1705-3}{{\em
  Eur. Phys. J.} {\bf C71} (2011) 1705}
  [\href{http://arXiv.org/abs/1012.4408}{{\tt arXiv:1012.4408 [hep-ph]}}].

\bibitem{Rezaeian:2012ji}
A.~H. Rezaeian, M.~Siddikov, M.~Van~de Klundert and R.~Venugopalan, {\it
  {Analysis of combined HERA data in the Impact-Parameter dependent Saturation
  model}},  \href{http://dx.doi.org/10.1103/PhysRevD.87.034002}{{\em Phys.
  Rev.} {\bf D87} (2013) 034002} [\href{http://arXiv.org/abs/1212.2974}{{\tt
  arXiv:1212.2974}}].

\bibitem{Lappi:2013zma}
T.~Lappi and H.~M{\"a}ntysaari, {\it {Single inclusive particle production at
  high energy from HERA data to proton-nucleus collisions}},
  \href{http://dx.doi.org/10.1103/PhysRevD.88.114020}{{\em Phys. Rev.} {\bf
  D88} (2013) 114020} [\href{http://arXiv.org/abs/1309.6963}{{\tt
  arXiv:1309.6963 [hep-ph]}}].

\bibitem{Tribedy:2010ab}
P.~Tribedy and R.~Venugopalan, {\it {Saturation models of HERA DIS data and
  inclusive hadron distributions in p+p collisions at the LHC}},
  \href{http://dx.doi.org/10.1016/j.nuclphysa.2011.04.008,
  10.1016/j.nuclphysa.2010.12.006}{{\em Nucl. Phys.} {\bf A850} (2011) 136}
  [\href{http://arXiv.org/abs/1011.1895}{{\tt arXiv:1011.1895 [hep-ph]}}].

\bibitem{Fujii:2013gxa}
H.~Fujii and K.~Watanabe, {\it {Heavy quark pair production in high energy pA
  collisions: Quarkonium}},
  \href{http://dx.doi.org/10.1016/j.nuclphysa.2013.06.011}{{\em Nucl. Phys.}
  {\bf A915} (2013) 1} [\href{http://arXiv.org/abs/1304.2221}{{\tt
  arXiv:1304.2221 [hep-ph]}}].

\bibitem{Ma:2014mri}
Y.-Q. Ma and R.~Venugopalan, {\it {Comprehensive Description of $J/\Psi$
  Production in Proton-Proton Collisions at Collider Energies}},
  \href{http://dx.doi.org/10.1103/PhysRevLett.113.192301}{{\em Phys. Rev.
  Lett.} {\bf 113} (2014) 192301} [\href{http://arXiv.org/abs/1408.4075}{{\tt
  arXiv:1408.4075 [hep-ph]}}].

\bibitem{Ducloue:2015gfa}
B.~Ducloué, T.~Lappi and H.~Mäntysaari, {\it {Forward $J/\psi$ production in
  proton-nucleus collisions at high energy}},
  \href{http://dx.doi.org/10.1103/PhysRevD.91.114005}{{\em Phys. Rev.} {\bf
  D91} (2015) 114005} [\href{http://arXiv.org/abs/1503.02789}{{\tt
  arXiv:1503.02789 [hep-ph]}}].

\bibitem{Ducloue:2016pqr}
B.~Ducloué, T.~Lappi and H.~Mäntysaari, {\it {Forward $J/\psi$ production at
  high energy: centrality dependence and mean transverse momentum}},
  \href{http://arXiv.org/abs/1605.05680}{{\tt arXiv:1605.05680 [hep-ph]}}.

\bibitem{Albacete:2010pg}
J.~L. Albacete and C.~Marquet, {\it {Azimuthal correlations of forward
  di-hadrons in d+Au collisions at RHIC in the Color Glass Condensate}},
  \href{http://dx.doi.org/10.1103/PhysRevLett.105.162301}{{\em Phys. Rev.
  Lett.} {\bf 105} (2010) 162301} [\href{http://arXiv.org/abs/1005.4065}{{\tt
  arXiv:1005.4065 [hep-ph]}}].

\bibitem{Stasto:2011ru}
A.~Stasto, B.-W. Xiao and F.~Yuan, {\it {Back-to-Back Correlations of
  Di-hadrons in dAu Collisions at RHIC}},
  \href{http://dx.doi.org/10.1016/j.physletb.2012.08.044}{{\em Phys. Lett.}
  {\bf B716} (2012) 430} [\href{http://arXiv.org/abs/1109.1817}{{\tt
  arXiv:1109.1817 [hep-ph]}}].

\bibitem{Lappi:2012nh}
T.~Lappi and H.~M{\"a}ntysaari, {\it {Forward dihadron correlations in
  deuteron-gold collisions with the Gaussian approximation of JIMWLK}},
  \href{http://dx.doi.org/10.1016/j.nuclphysa.2013.03.017}{{\em Nucl.Phys.}
  {\bf A908} (2013) 51} [\href{http://arXiv.org/abs/1209.2853}{{\tt
  arXiv:1209.2853 [hep-ph]}}].

\bibitem{Kovchegov:1999ji}
Y.~V. Kovchegov and E.~Levin, {\it {Diffractive dissociation including multiple
  pomeron exchanges in high parton density QCD}},
  \href{http://dx.doi.org/10.1016/S0550-3213(00)00125-5}{{\em Nucl. Phys.} {\bf
  B577} (2000) 221} [\href{http://arXiv.org/abs/hep-ph/9911523}{{\tt
  arXiv:hep-ph/9911523 [hep-ph]}}].

\bibitem{kuroda:2005by}
M.~Kuroda and D.~Schildknecht, {\it {$J/\Psi$ photo- and electroproduction, the
  saturation scale and the gluon structure function}},
  \href{http://dx.doi.org/10.1016/j.physletb.2006.05.070}{{\em Phys. Lett.}
  {\bf B638} (2006) 473} [\href{http://arXiv.org/abs/hep-ph/0507098}{{\tt
  arXiv:hep-ph/0507098 [hep-ph]}}].

\bibitem{Goncalves:2005yr}
V.~P. Goncalves and M.~V.~T. Machado, {\it {The QCD pomeron in ultraperipheral
  heavy ion collisions. IV. Photonuclear production of vector mesons}},
  \href{http://dx.doi.org/10.1140/epjc/s2005-02175-3}{{\em Eur. Phys. J.} {\bf
  C40} (2005) 519} [\href{http://arXiv.org/abs/hep-ph/0501099}{{\tt
  arXiv:hep-ph/0501099 [hep-ph]}}].

\bibitem{Kowalski:2008sa}
H.~Kowalski, T.~Lappi, C.~Marquet and R.~Venugopalan, {\it {Nuclear enhancement
  and suppression of diffractive structure functions at high energies}},
  \href{http://dx.doi.org/10.1103/PhysRevC.78.045201}{{\em Phys. Rev.} {\bf
  C78} (2008) 045201} [\href{http://arXiv.org/abs/0805.4071}{{\tt
  arXiv:0805.4071 [hep-ph]}}].

\bibitem{Caldwell:2009ke}
A.~Caldwell and H.~Kowalski, {\it {Investigating the gluonic structure of
  nuclei via $J/\Psi$ scattering}},
  \href{http://dx.doi.org/10.1103/PhysRevC.81.025203}{{\em Phys. Rev.} {\bf
  C81} (2010) 025203} [\href{http://arXiv.org/abs/0909.1254}{{\tt
  arXiv:0909.1254}}].

\bibitem{Lappi:2010dd}
T.~Lappi and H.~M{\"a}ntysaari, {\it {Incoherent diffractive
  $J/\Psi$-production in high energy nuclear DIS}},
  \href{http://dx.doi.org/10.1103/PhysRevC.83.065202}{{\em Phys. Rev.} {\bf
  C83} (2011) 065202} [\href{http://arXiv.org/abs/1011.1988}{{\tt
  arXiv:1011.1988 [hep-ph]}}].

\bibitem{Toll:2012mb}
T.~Toll and T.~Ullrich, {\it {Exclusive diffractive processes in electron-ion
  collisions}},  \href{http://dx.doi.org/10.1103/PhysRevC.87.024913}{{\em Phys.
  Rev.} {\bf C87} (2013) 024913} [\href{http://arXiv.org/abs/1211.3048}{{\tt
  arXiv:1211.3048 [hep-ph]}}].

\bibitem{Lappi:2013am}
T.~Lappi and H.~M{\"a}ntysaari, {\it {$J/\Psi$ production in ultraperipheral
  Pb+Pb and p+Pb collisions at LHC energies}},
  \href{http://dx.doi.org/10.1103/PhysRevC.87.032201}{{\em Phys. Rev.} {\bf
  C87} (2013) 032201} [\href{http://arXiv.org/abs/1301.4095}{{\tt
  arXiv:1301.4095 [hep-ph]}}].

\bibitem{Lappi:2014foa}
T.~Lappi, H.~Mäntysaari and R.~Venugopalan, {\it {Ballistic protons in
  incoherent exclusive vector meson production as a measure of rare parton
  fluctuations at an Electron-Ion Collider}},
  \href{http://dx.doi.org/10.1103/PhysRevLett.114.082301}{{\em Phys. Rev.
  Lett.} {\bf 114} (2015) 082301} [\href{http://arXiv.org/abs/1411.0887}{{\tt
  arXiv:1411.0887 [hep-ph]}}].

\bibitem{Dominguez:2008aa}
F.~Dominguez, C.~Marquet and B.~Wu, {\it {On multiple scatterings of mesons in
  hot and cold QCD matter}},
  \href{http://dx.doi.org/10.1016/j.nuclphysa.2009.03.008}{{\em Nucl. Phys.}
  {\bf A823} (2009) 99} [\href{http://arXiv.org/abs/0812.3878}{{\tt
  arXiv:0812.3878 [nucl-th]}}].

\bibitem{Marquet:2009vs}
C.~Marquet and B.~Wu, {\it {Exclusive vs. diffractive vector meson production
  in DIS at small x or off nuclei}},
  \href{http://arXiv.org/abs/0908.4180}{{\tt arXiv:0908.4180 [hep-ph]}}.

\bibitem{Boer:2011fh}
D.~Boer, M.~Diehl, R.~Milner, R.~Venugopalan, W.~Vogelsang {\em et.~al.}, {\it
  {Gluons and the quark sea at high energies: Distributions, polarization,
  tomography}},  \href{http://arXiv.org/abs/1108.1713}{{\tt arXiv:1108.1713
  [nucl-th]}}.

\bibitem{Accardi:2012qut}
A.~Accardi, J.~Albacete, M.~Anselmino, N.~Armesto, E.~Aschenauer {\em et.~al.},
  {\it {Electron Ion Collider: The Next QCD Frontier - Understanding the glue
  that binds us all}},  \href{http://arXiv.org/abs/1212.1701}{{\tt
  arXiv:1212.1701 [nucl-ex]}}.

\bibitem{Abelev:2012ba}
{\bf ALICE} collaboration, B.~Abelev {\em et.~al.}, {\it {Coherent $J/\psi$
  photoproduction in ultra-peripheral Pb-Pb collisions at $\sqrt{s_{NN}} =
  2.76$ TeV}},  \href{http://dx.doi.org/10.1016/j.physletb.2012.11.059}{{\em
  Phys. Lett.} {\bf B718} (2013) 1273}
  [\href{http://arXiv.org/abs/1209.3715}{{\tt arXiv:1209.3715 [nucl-ex]}}].

\bibitem{Abbas:2013oua}
{\bf ALICE} collaboration, E.~Abbas {\em et.~al.}, {\it {Charmonium and
  $e^+e^-$ pair photoproduction at mid-rapidity in ultra-peripheral Pb-Pb
  collisions at $\sqrt{s_{NN}}$ = 2.76 TeV}},
  \href{http://dx.doi.org/10.1140/epjc/s10052-013-2617-1}{{\em Eur. Phys. J.}
  {\bf C73} (2013) 2617} [\href{http://arXiv.org/abs/1305.1467}{{\tt
  arXiv:1305.1467 [nucl-ex]}}].

\bibitem{Khachatryan:2016qhq}
{\bf CMS} collaboration, V.~Khachatryan {\em et.~al.}, {\it {Coherent $J/\Psi$
  photoproduction in ultra-peripheral PbPb collisions at
  $\sqrt{s_{\mathrm{NN}}} = 2.76 $TeV with the CMS experiment}},
  \href{http://arXiv.org/abs/1605.06966}{{\tt arXiv:1605.06966 [nucl-ex]}}.

\bibitem{ATLAS:2016vdy}
{\bf ATLAS} collaboration, ``\emph{Measurement of high-mass dimuon pairs from
  ultraperipheral lead-lead collisions at $\sqrt{s_{\mathrm{NN}}}=5.02$ TeV
  with the ATLAS detector at the LHC}.''
\newblock ATLAS-CONF-2016-025.

\bibitem{TheALICE:2014dwa}
{\bf ALICE} collaboration, B.~B. Abelev {\em et.~al.}, {\it {Exclusive $J/\Psi$
  photoproduction off protons in ultra-peripheral p-Pb collisions at
  $\sqrt{s_{\rm NN}}=5.02$ TeV}},
  \href{http://dx.doi.org/10.1103/PhysRevLett.113.232504}{{\em Phys. Rev.
  Lett.} {\bf 113} (2014) 232504} [\href{http://arXiv.org/abs/1406.7819}{{\tt
  arXiv:1406.7819 [nucl-ex]}}].

\bibitem{CMS:2016nct}
{\bf CMS} collaboration, ``\emph{Measurement of exclusive $\Upsilon$
  photoproduction in pPb collisions at $\sqrt{s_{\mathrm{NN}}} =
  5.02~\mathrm{TeV}$}.''
\newblock CMS-PAS-FSQ-13-009.

\bibitem{Khachatryan:2010gv}
{\bf CMS} collaboration, V.~Khachatryan {\em et.~al.}, {\it {Observation of
  Long-Range Near-Side Angular Correlations in Proton-Proton Collisions at the
  LHC}},  \href{http://dx.doi.org/10.1007/JHEP09(2010)091}{{\em JHEP} {\bf 09}
  (2010) 091} [\href{http://arXiv.org/abs/1009.4122}{{\tt arXiv:1009.4122
  [hep-ex]}}].

\bibitem{Abelev:2012ola}
{\bf ALICE} collaboration, B.~Abelev {\em et.~al.}, {\it {Long-range angular
  correlations on the near and away side in $p$-Pb collisions at
  $\sqrt{s_{NN}}=5.02$ TeV}},
  \href{http://dx.doi.org/10.1016/j.physletb.2013.01.012}{{\em Phys. Lett.}
  {\bf B719} (2013) 29} [\href{http://arXiv.org/abs/1212.2001}{{\tt
  arXiv:1212.2001 [nucl-ex]}}].

\bibitem{Aad:2012gla}
{\bf ATLAS} collaboration, G.~Aad {\em et.~al.}, {\it {Observation of
  Associated Near-Side and Away-Side Long-Range Correlations in
  $\sqrt{s_{NN}}=5.02$  TeV Proton-Lead Collisions with the ATLAS
  Detector}},  \href{http://dx.doi.org/10.1103/PhysRevLett.110.182302}{{\em
  Phys. Rev. Lett.} {\bf 110} (2013) 182302}
  [\href{http://arXiv.org/abs/1212.5198}{{\tt arXiv:1212.5198 [hep-ex]}}].

\bibitem{Adare:2013piz}
{\bf PHENIX} collaboration, A.~Adare {\em et.~al.}, {\it {Quadrupole Anisotropy
  in Dihadron Azimuthal Correlations in Central $d+Au$ Collisions at
  $\sqrt{s_{_{NN}}}=200$ GeV}},
  \href{http://dx.doi.org/10.1103/PhysRevLett.111.212301}{{\em Phys. Rev.
  Lett.} {\bf 111} (2013) 212301} [\href{http://arXiv.org/abs/1303.1794}{{\tt
  arXiv:1303.1794 [nucl-ex]}}].

\bibitem{Dusling:2015gta}
K.~Dusling, W.~Li and B.~Schenke, {\it {Novel collective phenomena in
  high-energy proton–proton and proton–nucleus collisions}},
  \href{http://dx.doi.org/10.1142/S0218301316300022}{{\em Int. J. Mod. Phys.}
  {\bf E25} (2016) 1630002} [\href{http://arXiv.org/abs/1509.07939}{{\tt
  arXiv:1509.07939 [nucl-ex]}}].

\bibitem{Gale:2013da}
C.~Gale, S.~Jeon and B.~Schenke, {\it {Hydrodynamic Modeling of Heavy-Ion
  Collisions}},  \href{http://dx.doi.org/10.1142/S0217751X13400113}{{\em Int.
  J. Mod. Phys.} {\bf A28} (2013) 1340011}
  [\href{http://arXiv.org/abs/1301.5893}{{\tt arXiv:1301.5893 [nucl-th]}}].

\bibitem{Schenke:2014zha}
B.~Schenke and R.~Venugopalan, {\it {Eccentric protons? Sensitivity of flow to
  system size and shape in p+p, p+Pb and Pb+Pb collisions}},
  \href{http://dx.doi.org/10.1103/PhysRevLett.113.102301}{{\em Phys. Rev.
  Lett.} {\bf 113} (2014) 102301} [\href{http://arXiv.org/abs/1405.3605}{{\tt
  arXiv:1405.3605 [nucl-th]}}].

\bibitem{Deng:2011at}
W.-T. Deng, Z.~Xu and C.~Greiner, {\it {Elliptic and Triangular Flow and their
  Correlation in Ultrarelativistic High Multiplicity Proton Proton Collisions
  at 14 TeV}},  \href{http://dx.doi.org/10.1016/j.physletb.2012.04.010}{{\em
  Phys. Lett.} {\bf B711} (2012) 301}
  [\href{http://arXiv.org/abs/1112.0470}{{\tt arXiv:1112.0470 [hep-ph]}}].

\bibitem{Coleman-Smith:2013rla}
C.~E. Coleman-Smith and B.~Müller, {\it {Mapping the proton’s fluctuating
  size and shape}},  \href{http://dx.doi.org/10.1103/PhysRevD.89.025019}{{\em
  Phys. Rev.} {\bf D89} (2014) 025019}
  [\href{http://arXiv.org/abs/1307.5911}{{\tt arXiv:1307.5911 [hep-ph]}}].

\bibitem{Alvioli:2014eda}
M.~Alvioli, B.~A. Cole, L.~Frankfurt, D.~V. Perepelitsa and M.~Strikman, {\it
  {Evidence for $x$-dependent proton color fluctuations in pA collisions at the
  CERN Large Hadron Collider}},
  \href{http://dx.doi.org/10.1103/PhysRevC.93.011902}{{\em Phys. Rev.} {\bf
  C93} (2016)~no.~1 011902} [\href{http://arXiv.org/abs/1409.7381}{{\tt
  arXiv:1409.7381 [hep-ph]}}].

\bibitem{Aad:2015zza}
{\bf ATLAS} collaboration, G.~Aad {\em et.~al.}, {\it {Measurement of the
  centrality dependence of the charged-particle pseudorapidity distribution in
  proton–lead collisions at $\sqrt{s_{_\text {NN}}} = 5.02$ TeV with the
  ATLAS detector}},
  \href{http://dx.doi.org/10.1140/epjc/s10052-016-4002-3}{{\em Eur. Phys. J.}
  {\bf C76} (2016)~no.~4 199} [\href{http://arXiv.org/abs/1508.00848}{{\tt
  arXiv:1508.00848 [hep-ex]}}].

\bibitem{Welsh:2016siu}
K.~Welsh, J.~Singer and U.~W. Heinz, {\it {Initial state fluctuations in
  collisions between light and heavy ions}},
  \href{http://arXiv.org/abs/1605.09418}{{\tt arXiv:1605.09418 [nucl-th]}}.

\bibitem{Albacete:2016pmp}
J.~L. Albacete and A.~Soto-Ontoso, {\it {Hot spots and the hollowness of
  proton-proton interactions at high energies}},
  \href{http://arXiv.org/abs/1605.09176}{{\tt arXiv:1605.09176 [hep-ph]}}.

\bibitem{Good:1960ba}
M.~L. Good and W.~D. Walker, {\it {Diffraction disssociation of beam
  particles}},  \href{http://dx.doi.org/10.1103/PhysRev.120.1857}{{\em Phys.
  Rev.} {\bf 120} (1960) 1857}.

\bibitem{Miettinen:1978jb}
H.~I. Miettinen and J.~Pumplin, {\it {Diffraction Scattering and the Parton
  Structure of Hadrons}},
  \href{http://dx.doi.org/10.1103/PhysRevD.18.1696}{{\em Phys. Rev.} {\bf D18}
  (1978) 1696}.

\bibitem{Frankfurt:1993qi}
L.~Frankfurt, G.~A. Miller and M.~Strikman, {\it {Evidence for color
  fluctuations in hadrons from coherent nuclear diffraction}},
  \href{http://dx.doi.org/10.1103/PhysRevLett.71.2859}{{\em Phys. Rev. Lett.}
  {\bf 71} (1993) 2859} [\href{http://arXiv.org/abs/hep-ph/9309285}{{\tt
  arXiv:hep-ph/9309285 [hep-ph]}}].

\bibitem{Frankfurt:2008vi}
L.~Frankfurt, M.~Strikman, D.~Treleani and C.~Weiss, {\it {Evidence for color
  fluctuations in the nucleon in high-energy scattering}},
  \href{http://dx.doi.org/10.1103/PhysRevLett.101.202003}{{\em Phys. Rev.
  Lett.} {\bf 101} (2008) 202003} [\href{http://arXiv.org/abs/0808.0182}{{\tt
  arXiv:0808.0182 [hep-ph]}}].

\bibitem{Barone:2002cv}
V.~Barone and E.~Predazzi, {\em {High-Energy Particle Diffraction}}, vol.~565
  of {\em Texts and Monographs in Physics}.
\newblock Springer-Verlag, Berlin Heidelberg, 2002.

\bibitem{Kowalski:2006hc}
H.~Kowalski, L.~Motyka and G.~Watt, {\it {Exclusive diffractive processes at
  HERA within the dipole picture}},
  \href{http://dx.doi.org/10.1103/PhysRevD.74.074016}{{\em Phys. Rev.} {\bf
  D74} (2006) 074016} [\href{http://arXiv.org/abs/hep-ph/0606272}{{\tt
  arXiv:hep-ph/0606272}}].

\bibitem{Kovchegov:1999kx}
Y.~V. Kovchegov and L.~D. McLerran, {\it {Diffractive structure function in a
  quasiclassical approximation}},
  \href{http://dx.doi.org/10.1103/PhysRevD.62.019901,
  10.1103/PhysRevD.60.054025}{{\em Phys. Rev.} {\bf D60} (1999) 054025}
  [\href{http://arXiv.org/abs/hep-ph/9903246}{{\tt arXiv:hep-ph/9903246
  [hep-ph]}}].
\newblock [Erratum: Phys. Rev.D62,019901(2000)].

\bibitem{Kovner:2001vi}
A.~Kovner and U.~A. Wiedemann, {\it {Eikonal evolution and gluon radiation}},
  \href{http://dx.doi.org/10.1103/PhysRevD.64.114002}{{\em Phys. Rev.} {\bf
  D64} (2001) 114002} [\href{http://arXiv.org/abs/hep-ph/0106240}{{\tt
  arXiv:hep-ph/0106240 [hep-ph]}}].

\bibitem{Buchmuller:1998jv}
W.~Buchmuller, T.~Gehrmann and A.~Hebecker, {\it {Inclusive and diffractive
  structure functions at small x}},
  \href{http://dx.doi.org/10.1016/S0550-3213(98)00682-8}{{\em Nucl. Phys.} {\bf
  B537} (1999) 477} [\href{http://arXiv.org/abs/hep-ph/9808454}{{\tt
  arXiv:hep-ph/9808454 [hep-ph]}}].

\bibitem{Kovchegov:2012mbw}
Y.~V. Kovchegov and E.~Levin, {\em {Quantum chromodynamics at high energy}}.
\newblock Cambridge University Press, 2012.

\bibitem{JalilianMarian:1996xn}
J.~Jalilian-Marian, A.~Kovner, L.~D. McLerran and H.~Weigert, {\it {The
  Intrinsic glue distribution at very small $x$}},
  \href{http://dx.doi.org/10.1103/PhysRevD.55.5414}{{\em Phys. Rev.} {\bf D55}
  (1997) 5414} [\href{http://arXiv.org/abs/hep-ph/9606337}{{\tt
  arXiv:hep-ph/9606337 [hep-ph]}}].

\bibitem{JalilianMarian:1997jx}
J.~Jalilian-Marian, A.~Kovner, A.~Leonidov and H.~Weigert, {\it {The BFKL
  equation from the Wilson renormalization group}},
  \href{http://dx.doi.org/10.1016/S0550-3213(97)00440-9}{{\em Nucl. Phys.} {\bf
  B504} (1997) 415} [\href{http://arXiv.org/abs/hep-ph/9701284}{{\tt
  arXiv:hep-ph/9701284 [hep-ph]}}].

\bibitem{JalilianMarian:1997gr}
J.~Jalilian-Marian, A.~Kovner, A.~Leonidov and H.~Weigert, {\it {The Wilson
  renormalization group for low $x$ physics: Towards the high density regime}},
   \href{http://dx.doi.org/10.1103/PhysRevD.59.014014}{{\em Phys. Rev.} {\bf
  D59} (1998) 014014} [\href{http://arXiv.org/abs/hep-ph/9706377}{{\tt
  arXiv:hep-ph/9706377 [hep-ph]}}].

\bibitem{Iancu:2001md}
E.~Iancu and L.~D. McLerran, {\it {Saturation and universality in QCD at small
  $x$}},  \href{http://dx.doi.org/10.1016/S0370-2693(01)00526-3}{{\em Phys.
  Lett.} {\bf B510} (2001) 145}
  [\href{http://arXiv.org/abs/hep-ph/0103032}{{\tt arXiv:hep-ph/0103032
  [hep-ph]}}].

\bibitem{Balitsky:1995ub}
I.~Balitsky, {\it {Operator expansion for high-energy scattering}},
  \href{http://dx.doi.org/10.1016/0550-3213(95)00638-9}{{\em Nucl. Phys.} {\bf
  B463} (1996) 99} [\href{http://arXiv.org/abs/hep-ph/9509348}{{\tt
  arXiv:hep-ph/9509348}}].

\bibitem{Kovchegov:1999yj}
Y.~V. Kovchegov, {\it {Small-$x$ $F_2$ structure function of a nucleus
  including multiple pomeron exchanges}},
  \href{http://dx.doi.org/10.1103/PhysRevD.60.034008}{{\em Phys. Rev.} {\bf
  D60} (1999) 034008} [\href{http://arXiv.org/abs/hep-ph/9901281}{{\tt
  arXiv:hep-ph/9901281 [hep-ph]}}].

\bibitem{Kowalski:2003hm}
H.~Kowalski and D.~Teaney, {\it {An Impact parameter dipole saturation model}},
   \href{http://dx.doi.org/10.1103/PhysRevD.68.114005}{{\em Phys. Rev.} {\bf
  D68} (2003) 114005} [\href{http://arXiv.org/abs/hep-ph/0304189}{{\tt
  arXiv:hep-ph/0304189 [hep-ph]}}].

\bibitem{GolecBiernat:2003ym}
K.~J. Golec-Biernat and A.~Stasto, {\it {On solutions of the Balitsky-Kovchegov
  equation with impact parameter}},
  \href{http://dx.doi.org/10.1016/j.nuclphysb.2003.07.011}{{\em Nucl. Phys.}
  {\bf B668} (2003) 345} [\href{http://arXiv.org/abs/hep-ph/0306279}{{\tt
  arXiv:hep-ph/0306279 [hep-ph]}}].

\bibitem{Schlichting:2014ipa}
S.~Schlichting and B.~Schenke, {\it {The shape of the proton at high
  energies}},  \href{http://dx.doi.org/10.1016/j.physletb.2014.10.068}{{\em
  Phys. Lett.} {\bf B739} (2014) 313}
  [\href{http://arXiv.org/abs/1407.8458}{{\tt arXiv:1407.8458 [hep-ph]}}].

\bibitem{Schenke:2012wb}
B.~Schenke, P.~Tribedy and R.~Venugopalan, {\it {Fluctuating Glasma initial
  conditions and flow in heavy ion collisions}},
  \href{http://dx.doi.org/10.1103/PhysRevLett.108.252301}{{\em Phys. Rev.
  Lett.} {\bf 108} (2012) 252301} [\href{http://arXiv.org/abs/1202.6646}{{\tt
  arXiv:1202.6646 [nucl-th]}}].

\bibitem{Schenke:2013dpa}
B.~Schenke, P.~Tribedy and R.~Venugopalan, {\it {Multiplicity distributions in
  p+p, p+A and A+A collisions from Yang-Mills dynamics}},
  \href{http://dx.doi.org/10.1103/PhysRevC.89.024901}{{\em Phys. Rev.} {\bf
  C89} (2014) 024901} [\href{http://arXiv.org/abs/1311.3636}{{\tt
  arXiv:1311.3636 [hep-ph]}}].

\bibitem{Bartels:2002cj}
J.~Bartels, K.~J. Golec-Biernat and H.~Kowalski, {\it {A modification of the
  saturation model: DGLAP evolution}},
  \href{http://dx.doi.org/10.1103/PhysRevD.66.014001}{{\em Phys. Rev.} {\bf
  D66} (2002) 014001} [\href{http://arXiv.org/abs/hep-ph/0203258}{{\tt
  arXiv:hep-ph/0203258 [hep-ph]}}].

\bibitem{Gotsman:2001ne}
E.~Gotsman, E.~Levin, U.~Maor and E.~Naftali, {\it {Momentum transfer
  dependence of the differential cross-section for $J/\Psi$ production}},
  \href{http://dx.doi.org/10.1016/S0370-2693(02)01542-3}{{\em Phys. Lett.} {\bf
  B532} (2002) 37} [\href{http://arXiv.org/abs/hep-ph/0110256}{{\tt
  arXiv:hep-ph/0110256 [hep-ph]}}].

\bibitem{Lappi:2007ku}
T.~Lappi, {\it {Wilson line correlator in the MV model: Relating the glasma to
  deep inelastic scattering}},
  \href{http://dx.doi.org/10.1140/epjc/s10052-008-0588-4}{{\em Eur. Phys. J.}
  {\bf C55} (2008) 285} [\href{http://arXiv.org/abs/0711.3039}{{\tt
  arXiv:0711.3039 [hep-ph]}}].

\bibitem{Schenke:2012fw}
B.~Schenke, P.~Tribedy and R.~Venugopalan, {\it {Event-by-event gluon
  multiplicity, energy density, and eccentricities in ultrarelativistic
  heavy-ion collisions}},
  \href{http://dx.doi.org/10.1103/PhysRevC.86.034908}{{\em Phys. Rev.} {\bf
  C86} (2012) 034908} [\href{http://arXiv.org/abs/1206.6805}{{\tt
  arXiv:1206.6805 [hep-ph]}}].

\bibitem{Aktas:2005xu}
{\bf H1} collaboration, A.~Aktas {\em et.~al.}, {\it {Elastic $J/\Psi$
  production at HERA}},
  \href{http://dx.doi.org/10.1140/epjc/s2006-02519-5}{{\em Eur. Phys. J.} {\bf
  C46} (2006) 585} [\href{http://arXiv.org/abs/hep-ex/0510016}{{\tt
  arXiv:hep-ex/0510016}}].

\bibitem{Martin:1997wy}
A.~D. Martin and M.~G. Ryskin, {\it {The effect of off diagonal parton
  distributions in diffractive vector meson electroproduction}},
  \href{http://dx.doi.org/10.1103/PhysRevD.57.6692}{{\em Phys. Rev.} {\bf D57}
  (1998) 6692} [\href{http://arXiv.org/abs/hep-ph/9711371}{{\tt
  arXiv:hep-ph/9711371 [hep-ph]}}].

\bibitem{Shuvaev:1999ce}
A.~G. Shuvaev, K.~J. Golec-Biernat, A.~D. Martin and M.~G. Ryskin, {\it
  {Off-diagonal distributions fixed by diagonal partons at small $x$ and
  $\xi$}},  \href{http://dx.doi.org/10.1103/PhysRevD.60.014015}{{\em Phys.
  Rev.} {\bf D60} (1999) 014015}
  [\href{http://arXiv.org/abs/hep-ph/9902410}{{\tt arXiv:hep-ph/9902410}}].

\bibitem{Martin:1999wb}
A.~D. Martin, M.~Ryskin and T.~Teubner, {\it {$Q^2$ dependence of diffractive
  vector meson electroproduction}},
  \href{http://dx.doi.org/10.1103/PhysRevD.62.014022}{{\em Phys. Rev.} {\bf
  D62} (2000) 014022} [\href{http://arXiv.org/abs/hep-ph/9912551}{{\tt
  arXiv:hep-ph/9912551 [hep-ph]}}].

\bibitem{RevModPhys.28.214}
R.~Hofstadter, {\it Electron scattering and nuclear structure},
  \href{http://dx.doi.org/10.1103/RevModPhys.28.214}{{\em Rev. Mod. Phys.} {\bf
  28} (Jul, 1956) 214}.

\bibitem{Bernauer:2010wm}
{\bf A1} collaboration, J.~C. Bernauer {\em et.~al.}, {\it {High-precision
  determination of the electric and magnetic form factors of the proton}},
  \href{http://dx.doi.org/10.1103/PhysRevLett.105.242001}{{\em Phys. Rev.
  Lett.} {\bf 105} (2010) 242001} [\href{http://arXiv.org/abs/1007.5076}{{\tt
  arXiv:1007.5076 [nucl-ex]}}].

\bibitem{Zhan:2011ji}
X.~Zhan {\em et.~al.}, {\it {High-Precision Measurement of the Proton Elastic
  Form Factor Ratio $\mu_pG_E/G_M$ at low $Q^2$}},
  \href{http://dx.doi.org/10.1016/j.physletb.2011.10.002}{{\em Phys. Lett.}
  {\bf B705} (2011) 59} [\href{http://arXiv.org/abs/1102.0318}{{\tt
  arXiv:1102.0318 [nucl-ex]}}].

\bibitem{Antognini417}
{\it Proton structure from the measurement of 2s-2p transition frequencies of
  muonic hydrogen},  \href{http://dx.doi.org/10.1126/science.1230016}{{\em
  Science} {\bf 339} (2013)~no.~6118 417}.

\bibitem{Friar:2003zg}
J.~L. Friar and I.~Sick, {\it {Zemach moments for hydrogen and deuterium}},
  \href{http://dx.doi.org/10.1016/j.physletb.2003.11.018}{{\em Phys. Lett.}
  {\bf B579} (2004) 285} [\href{http://arXiv.org/abs/nucl-th/0310043}{{\tt
  arXiv:nucl-th/0310043 [nucl-th]}}].

\bibitem{Distler:2010zq}
M.~O. Distler, J.~C. Bernauer and T.~Walcher, {\it {The RMS Charge Radius of
  the Proton and Zemach Moments}},
  \href{http://dx.doi.org/10.1016/j.physletb.2010.12.067}{{\em Phys. Lett.}
  {\bf B696} (2011) 343} [\href{http://arXiv.org/abs/1011.1861}{{\tt
  arXiv:1011.1861 [nucl-th]}}].

\bibitem{Meissner:1987ge}
U.~G. Meissner, {\it {Low-Energy Hadron Physics from Effective Chiral
  Lagrangians with Vector Mesons}},
  \href{http://dx.doi.org/10.1016/0370-1573(88)90090-7}{{\em Phys. Rept.} {\bf
  161} (1988) 213}.

\bibitem{Kawasaki:1998gn}
M.~Kawasaki, T.~Maehara and M.~Yonezawa, {\it {Gluonic radius of the proton and
  high-energy $\bar{p} p$ scattering}},
  \href{http://dx.doi.org/10.1103/PhysRevD.57.1822}{{\em Phys. Rev.} {\bf D57}
  (1998) 1822}.

\bibitem{Caldwell:2010zza}
A.~Caldwell and H.~Kowalski, {\it {Investigating the gluonic structure of
  nuclei via $J/\Psi$ scattering}},  in {\em Phys. Rev.\/}
  \cite{Caldwell:2009ke}, p.~025203,
  [\href{http://arXiv.org/abs/0909.1254}{{\tt arXiv:0909.1254}}].

\bibitem{Bissey:2006bz}
F.~Bissey, F.-G. Cao, A.~R. Kitson, A.~I. Signal, D.~B. Leinweber, B.~G.
  Lasscock and A.~G. Williams, {\it {Gluon flux-tube distribution and linear
  confinement in baryons}},
  \href{http://dx.doi.org/10.1103/PhysRevD.76.114512}{{\em Phys. Rev.} {\bf
  D76} (2007) 114512} [\href{http://arXiv.org/abs/hep-lat/0606016}{{\tt
  arXiv:hep-lat/0606016 [hep-lat]}}].

\bibitem{Alexa:2013xxa}
{\bf H1} collaboration, C.~Alexa {\em et.~al.}, {\it {Elastic and
  Proton-Dissociative Photoproduction of J$/\Psi$ Mesons at HERA}},
  \href{http://dx.doi.org/10.1140/epjc/s10052-013-2466-y}{{\em Eur. Phys. J.}
  {\bf C73} (2013)~no.~6 2466} [\href{http://arXiv.org/abs/1304.5162}{{\tt
  arXiv:1304.5162 [hep-ex]}}].

\bibitem{McLerran:2015qxa}
L.~McLerran and P.~Tribedy, {\it {Intrinsic Fluctuations of the Proton
  Saturation Momentum Scale in High Multiplicity p+p Collisions}},
  \href{http://dx.doi.org/10.1016/j.nuclphysa.2015.10.008}{{\em Nucl. Phys.}
  {\bf A945} (2016) 216} [\href{http://arXiv.org/abs/1508.03292}{{\tt
  arXiv:1508.03292 [hep-ph]}}].

\bibitem{Bzdak:2015eii}
A.~Bzdak and K.~Dusling, {\it {Probing proton fluctuations with asymmetric
  rapidity correlations}},
  \href{http://dx.doi.org/10.1103/PhysRevC.93.031901}{{\em Phys. Rev.} {\bf
  C93} (2016)~no.~3 031901} [\href{http://arXiv.org/abs/1511.03620}{{\tt
  arXiv:1511.03620 [hep-ph]}}].

\bibitem{Dumitru:2012yr}
A.~Dumitru and Y.~Nara, {\it {KNO scaling of fluctuations in pp and pA, and
  eccentricities in heavy-ion collisions}},
  \href{http://dx.doi.org/10.1103/PhysRevC.85.034907}{{\em Phys. Rev.} {\bf
  C85} (2012) 034907} [\href{http://arXiv.org/abs/1201.6382}{{\tt
  arXiv:1201.6382 [nucl-th]}}].

\bibitem{Chekanov:2002xi}
{\bf ZEUS} collaboration, S.~Chekanov {\em et.~al.}, {\it {Exclusive
  photoproduction of $J/\Psi$ mesons at HERA}},
  \href{http://dx.doi.org/10.1007/s10052-002-0953-7}{{\em Eur. Phys. J.} {\bf
  C24} (2002) 345} [\href{http://arXiv.org/abs/hep-ex/0201043}{{\tt
  arXiv:hep-ex/0201043 [hep-ex]}}].

\bibitem{Chekanov:2002rm}
{\bf ZEUS} collaboration, S.~Chekanov {\em et.~al.}, {\it {Measurement of
  proton dissociative diffractive photoproduction of vector mesons at large
  momentum transfer at HERA}},
  \href{http://dx.doi.org/10.1140/epjc/s2002-01079-0}{{\em Eur. Phys. J.} {\bf
  C26} (2003) 389} [\href{http://arXiv.org/abs/hep-ex/0205081}{{\tt
  arXiv:hep-ex/0205081 [hep-ex]}}].

\bibitem{Aktas:2003zi}
{\bf H1} collaboration, A.~Aktas {\em et.~al.}, {\it {Diffractive
  photoproduction of $J/\Psi$ mesons with large momentum transfer at HERA}},
  \href{http://dx.doi.org/10.1016/j.physletb.2003.06.056}{{\em Phys. Lett.}
  {\bf B568} (2003) 205} [\href{http://arXiv.org/abs/hep-ex/0306013}{{\tt
  arXiv:hep-ex/0306013 [hep-ex]}}].

\bibitem{Aid:1996ee}
{\bf H1} collaboration, S.~Aid {\em et.~al.}, {\it {Elastic Electroproduction
  of $\rho$ and $J/\Psi$ Mesons at large $Q^2$ at HERA}},
  \href{http://dx.doi.org/10.1016/0550-3213(96)00192-7}{{\em Nucl. Phys.} {\bf
  B468} (1996) 3} [\href{http://arXiv.org/abs/hep-ex/9602007}{{\tt
  arXiv:hep-ex/9602007 [hep-ex]}}].

\bibitem{Adloff:1999kg}
{\bf H1} collaboration, C.~Adloff {\em et.~al.}, {\it {Elastic
  electroproduction of $\rho$ mesons at HERA}},
  \href{http://dx.doi.org/10.1007/s100520050703}{{\em Eur. Phys. J.} {\bf C13}
  (2000) 371} [\href{http://arXiv.org/abs/hep-ex/9902019}{{\tt
  arXiv:hep-ex/9902019 [hep-ex]}}].

\bibitem{Breitweg:1999jy}
{\bf ZEUS} collaboration, J.~Breitweg {\em et.~al.}, {\it {Measurement of
  diffractive photoproduction of vector mesons at large momentum transfer at
  HERA}},  \href{http://dx.doi.org/10.1007/s100520050748}{{\em Eur. Phys. J.}
  {\bf C14} (2000) 213} [\href{http://arXiv.org/abs/hep-ex/9910038}{{\tt
  arXiv:hep-ex/9910038 [hep-ex]}}].

\bibitem{Chekanov:2005cqa}
{\bf ZEUS} collaboration, S.~Chekanov {\em et.~al.}, {\it {Exclusive
  electroproduction of phi mesons at HERA}},
  \href{http://dx.doi.org/10.1016/j.nuclphysb.2005.04.009}{{\em Nucl. Phys.}
  {\bf B718} (2005) 3} [\href{http://arXiv.org/abs/hep-ex/0504010}{{\tt
  arXiv:hep-ex/0504010 [hep-ex]}}].

\bibitem{Aaron:2009xp}
{\bf H1} collaboration, F.~Aaron {\em et.~al.}, {\it {Diffractive
  Electroproduction of $\rho$ and $\phi$ Mesons at HERA}},
  \href{http://dx.doi.org/10.1007/JHEP05(2010)032}{{\em JHEP} {\bf 1005} (2010)
  032} [\href{http://arXiv.org/abs/0910.5831}{{\tt arXiv:0910.5831 [hep-ex]}}].

\bibitem{Mitchell:2016jio}
J.~T. Mitchell, D.~V. Perepelitsa, M.~J. Tannenbaum and P.~W. Stankus, {\it
  {Tests of constituent-quark generation methods which maintain both the
  nucleon center of mass and the desired radial distribution in Monte Carlo
  Glauber models}},  \href{http://dx.doi.org/10.1103/PhysRevC.93.054910}{{\em
  Phys. Rev.} {\bf C93} (2016)~no.~5 054910}
  [\href{http://arXiv.org/abs/1603.08836}{{\tt arXiv:1603.08836 [nucl-ex]}}].

\bibitem{Flensburg:2012zy}
C.~Flensburg, G.~Gustafson and L.~Lönnblad, {\it {Exclusive final states in
  diffractive excitation}},
  \href{http://dx.doi.org/10.1007/JHEP12(2012)115}{{\em JHEP} {\bf 12} (2012)
  115} [\href{http://arXiv.org/abs/1210.2407}{{\tt arXiv:1210.2407 [hep-ph]}}].

\bibitem{Mueller:2014fba}
A.~H. Mueller and S.~Munier, {\it {On parton number fluctuations at various
  stages of the rapidity evolution}},
  \href{http://dx.doi.org/10.1016/j.physletb.2014.08.058}{{\em Phys. Lett.}
  {\bf B737} (2014) 303} [\href{http://arXiv.org/abs/1405.3131}{{\tt
  arXiv:1405.3131 [hep-ph]}}].

\end{thebibliography}\endgroup

\end{document}